\documentclass[10pt]{article}

\usepackage{amsmath}
\usepackage{amssymb}

\usepackage{graphicx}

\usepackage{cite}

\usepackage{hyperref}

\usepackage{lineno}

\usepackage{microtype}
\usepackage{color}
\DisableLigatures[f]{encoding = *, family = * }

\usepackage[labelfont=bf,labelsep=period,justification=raggedright]{caption}

\makeatletter
\renewcommand{\@biblabel}[1]{\quad#1.}
\makeatother

\date{}

\begin{document}

% Title must be 150 characters or less
\begin{flushleft}
{\Large
	\textbf{Self-organization of microcircuits in networks of spiking neurons with plastic synapses}
}
% Insert Author names, affiliations and corresponding author email.
\\
Gabriel Koch Ocker$^{1,3}$, 
Ashok Litwin-Kumar$^{2,3,4}$, 
Brent Doiron$^{2,3 \ast}$
\\
\bf{1:} Department of Neuroscience, University of Pittsburgh, Pittsburgh, PA, United States of America
\\
\bf{2:} Department of Mathematics, University of Pittsburgh, Pittsburgh, PA, United States of America
\\
\bf{3:} Center for the Neural Basis of Cognition, University of Pittsburgh and Carnegie Mellon University, Pittsburgh, PA, United States of America
\\
\bf{4:} Center for Theoretical Neuroscience, Columbia University, New York, NY, United States of America

$\ast$ E-mail: bdoiron@pitt.edu
\end{flushleft}

% Please keep the abstract between 250 and 300 words
\section*{Abstract}
\paragraph{}
The synaptic connectivity of cortical networks features an overrepresentation of certain wiring motifs compared to simple random-network models. This structure is shaped, in part, by synaptic plasticity that promotes or suppresses connections between neurons depending on their spiking activity.  Frequently, theoretical studies focus on how feedforward inputs drive plasticity to create this network structure.  We study the complementary scenario of self-organized structure in a recurrent network, with spike timing-dependent plasticity driven by spontaneous dynamics.  We develop a self-consistent theory that describes the evolution of network structure by combining fast spiking covariance with a fast-slow theory for synaptic weight dynamics.  Through a finite-size expansion of network dynamics, we obtain a low-dimensional set of nonlinear differential equations for the evolution of two-synapse connectivity motifs.  With this theory in hand, we explore how the form of the plasticity rule drives the evolution of microcircuits in cortical networks.  When potentiation and depression are in approximate balance, synaptic dynamics depend on the frequency of weighted divergent, convergent, and chain motifs.  For additive, Hebbian STDP, these motif interactions create instabilities in synaptic dynamics that either promote or suppress the initial network structure. Our work provides a consistent theoretical framework for studying how spiking activity in recurrent networks interacts with synaptic plasticity to determine network structure.      

% Please keep the Author Summary between 150 and 200 words
% Use first person. PLOS ONE authors please skip this step. 
% Author Summary not valid for PLOS ONE submissions.   
\section*{Author Summary}
\paragraph{}
The connectivity of mammalian brains exhibits structure at a wide variety of spatial scales, from the broad (which brain areas connect to which) to the extremely fine (where synapses from different inputs lie on the morphology of individual neurons).  Recent experimental work in the neocortex has highlighted structure at the level of microcircuits: different patterns of connectivity between small groups of neurons are either more or less abundant than would be expected by chance.  A central question in systems neuroscience is how this structure emerges.  Attempts to answer this question are confounded by the known mutual interaction of network structure and spiking activity.  Indeed, synaptic connections influence spiking statistics, while individual synapses are highly plastic and become stronger or weaker depending on the activity of the pre- and postsynaptic neurons.  We present a self-consistent theory for how activity-dependent synaptic plasticity leads to the emergence of neuronal microcircuits and use it to show how the form of the plasticity rule can govern the promotion or suppression of different connectivity patterns.  Our work provides a foundation for understanding how cortical circuits, and not just individual synapses, are malleable in response to inputs both external and internal to a network.

\section*{Introduction}
%\paragraph{}
%In the brain, neurons are connected in complex networks.  Individual neurons undergo stochastic dynamics and influence the statistics of each others' activity through synapses.  Furthermore, the structure of neuronal networks is itself dynamic; synapses can be created or lost via structural plasticity, and their strengths can change on short and long timescales in a way that depends on the joint activity of the pre- and postsynaptic neurons.  This interaction of spiking dynamics on a network and plasticity of the network structure poses a central challenge for theories of neuronal activity.  

\paragraph{}
The wiring of neuronal networks exhibits structure across a broad range of spatial scales \cite{bullmore_complex_2009}.  For example, patterns of connectivity among small groups of cortical neurons are over- or under-represented compared to random networks \cite{markram_network_1997, perin_synaptic_2011, song_highly_2005, yoshimura_excitatory_2005}.  The prevalence of these motifs is related to a neuron's stimulus preferences and activity levels \cite{ko_functional_2011, yassin_embedded_2010}.  Motivated in part by these observations, there is a growing body of theoretical work that discusses how wiring structure dictates the coordinated spiking activity of cortical neurons in recurrent networks \cite{zhao_synchronization_2011, roxin_role_2011, litwin-kumar_slow_2012, gaiteri_interaction_2011, kriener_correlations_2009, pernice_how_2011, pernice_relevance_2013, trousdale_impact_2012, hu_motif_2013, helias_correlation_2014, hu_local_2014}.  

\paragraph{}
While neural architecture undoubtedly plays a strong role in determining neuronal activity, the reverse is also true.  Individual synapses can both potentiate (strengthen) and depress (weaken), and whether they do so depends on the relative timing of action potentials from the connected neurons \cite{bi_synaptic_1998, markram_regulation_1997}.  Such {\it spike timing-dependent plasticity} (STDP) has featured prominently in both experimental and theoretical studies of neural circuits \cite{abbott_synaptic_2000, caporale_spike_2008,markram_history_2011}.  Of particular interest, STDP provides a mechanism for Hebbian plasticity: synaptic potentiation occurs when a presynaptic neuron reliably drives spike responses from a postsynaptic neuron, while failure to recruit spiking results in synaptic depression \cite{song_competitive_2000}.  Hebbian plasticity provides a potential link between circuit structure and function by providing a mechanism for the formation of heavily wired assemblies of neurons, where assembly membership is associated with coordinated, elevated firing rates during a specific computation \cite{harris_neural_2005}.  Evidence supporting this idea, originally proposed by Hebb \cite{hebb_organization_1949}, has been found in both hippocampus \cite{buzsaki_neural_2010} and sensory cortex \cite{harris_cortical_2013}.    

\paragraph{}
Despite the promise of STDP to provide insight into the functional wiring of large neural circuits, many studies of STDP have concentrated on the plasticity of synaptic connections between just a single pair of pre- and postsynaptic neurons, often focusing on the distribution of individual synaptic weights \cite{song_competitive_2000, kempter_intrinsic_2001, babadi_intrinsic_2010, guetig_learning_2003, rubin_equilibrium_2001}.  %In those studies, when depression is slightly stronger compared to potentiation or postsynaptic spikes drive depression in addition to STDP, then a neuron's firing rate can stabilize.  
Other studies have shown that multiple temporally correlated inputs to a neuron will cooperate to potentiate, while uncorrelated inputs may depress \cite{kempter_hebbian_1999, song_competitive_2000, meffin_learning_2006, gilson_emergence_2009-1}. In this case STDP can generate feedforward circuits \cite{fiete_spike-time-dependent_2010}, which while important for the propagation of neural activity \cite{kumar_spiking_2010}, are unlike the recurrent structure of the neocortex.  Understanding the two-way interaction between plastic recurrent network structure and spiking activity is thus a central challenge for theories of synaptic plasticity.

\paragraph{}
Due to this challenge, many studies have resorted to large scale numerical simulations of cortical networks with plastic synapses \cite{izhikevich_spike-timing_2004, morrison_spike-timing-dependent_2007, babadi_pairwise_2013, litwin-kumar_formation_2014}.  While intuition for the development of circuit structure can be gained using this approach, without a governing theoretical framework it is often difficult to extract generalized principles.  Alternatively, mathematical analyses have been restricted to either small networks \cite{karbowski_synchrony_2002, babadi_pairwise_2013}, or have required the assumption that neurons fire as Poisson processes \cite{burkitt_spike-timing-dependent_2007, gilson_emergence_2009-2, gilson_emergence_2009-3, gilson_emergence_2010}.  These latter works assumed shared inputs from outside the network to be the only source of correlated spiking activity, neglecting covariance due to recurrent coupling.  Thus, there is a need for a coherent mathematical framework that captures how STDP drives self-organization of circuit structure in recurrent cortical networks.    
 
\paragraph{}
To this end, we construct a self-consistent theory for the coevolution of spiking statistics and synaptic weights in networks with STDP.  This theory makes use of a previously developed linear response framework for calculating joint spiking statistics \cite{doiron_oscillatory_2004, lindner_theory_2005, trousdale_impact_2012} and a separation of timescales between spiking covariance and synaptic plasticity \cite{kempter_hebbian_1999}.  Most previous studies of plasticity in recurrent networks have focused on how they can be trained to represent an external stimulus.  We focus on how spiking covariance generated by coupling within the network interacts with plasticity to shape network structure.  %Our theory can easily be extended to incorporate spatially or temporally correlated external inputs.  
We then use this high-dimensional theory to derive a low-dimensional, closed system for STDP of synaptic connectivity motifs in recurrent networks.  This reveals instabilities in the motif dynamics such that when potentiation and depression are approximately balanced, the dynamics are partitioned into regimes in which different motifs are promoted or suppressed depending on the initial network structure.  It also highlights the circumstances in which spike time covariations, in contrast to firing rates, drive STDP.  In total, we provide a consistent and general framework in which to study STDP in large recurrent networks.
%: spike time correlations require a balance between potentiation and depression.

% You may title this section "Methods" or "Models". 
% "Models" is not a valid title for PLoS ONE authors. However, PLoS ONE
% authors may use "Analysis" 

\section*{Results}
\paragraph{}
Our study is separated into two main sections.  The first presents a self-consistent theory for STDP in recurrent networks of model spiking neurons.  Using this theory, we accurately predict the co-evolution of spiking covariance on fast timescales with synaptic weights on slow timescales of a large network.  The second part leverages our theory to develop a low-dimensional dynamical system for the development of two-synapse motifs in the network structure.  We analyze this system and determine how the balance between synaptic potentiation and depression drives the emergence of microcircuits in recurrent networks.     

% We only support three levels of headings, please do not create a heading level below \subsubsection.
\subsection*{Spike train covariance determines synaptic plasticity}
%\paragraph{}
%Many experimental studies have investigated how spike timing-dependent plasticity controls the strength of individual synapses (CITATIONS).  Understanding how STDP controls the structure of neuronal networks is an important area of investigation if we are to understand how brains develop, learn and process information, and has been the subject of much theoretical investigation (CITATIONS).  Importantly, the processes underlying synaptic plasticity are ongoing during both spontaneous and stimulus-evoked activity.  Investigating how plasticity during spontaneous network activity is an important first step in understanding how networks learn to encode external stimuli.  Previous simulation studies have shown that synaptic plasticity can tie rise to rich spontaneous dynamics in neuronsÕ firing rates which serve to reinforce learned associations (CITATIONS).  However, a unified theory for how synaptic plasticity shapes both network structure and the statistics of spike timing has so far been lacking.

\paragraph{}
We begin by reviewing a well studied phenomenological model of spike timing-dependent plasticity (STDP) \cite{gerstner_neuronal_1996}, acting within a simple circuit of two reciprocally coupled neurons.  Consider a pair of pre- and postsynaptic spike times with time lag $s=t_{\mathrm{post}}-t_{\mathrm{pre}}$.  The evolution of the synaptic weight connecting presynaptic neuron $j$ to postsynaptic neuron $i$ obeys $\mathbf{W}_{ij} \to \mathbf{W}_{ij} + L(s)$, with the STDP rule $L(s)$ (Fig. 1A) being Hebbian:
\begin{equation} \label{stdprule}
L(s)=\begin{cases} \mathcal{H}(W^\mathrm{max}-\mathbf{W}_{ij})f_+e^{-\frac{\left| s\right|}{\tau_+}}, &\mathrm{if } s \geq 0 \\
\mathcal{H}(\mathbf{W}_{ij}) \left(-f_-\right)e^{-\frac{\left| s\right|}{\tau_-}}, &\mathrm{if } s \leq 0,
\end{cases}.
\end{equation}
Here $\mathcal{H}(x)=1$ if $x>0$ while $\mathcal{H}(x)=0$ if $x<0$, imposing bounds on the weights to prevent the magnitude of excitatory synapses from becoming negative or potentiating without bound (i.e. $0 \le \mathbf{W}_{ij} \le W^\mathrm{max}$).  The coefficients $f_\pm$ scale the amplitude of weight changes induced by individual pre-post spike pairs and $\tau_\pm$ determine how synchronous pre- and postsynaptic spikes must be to drive plasticity.  

\paragraph{}
The spike train from neuron $i$ is the point process $\mathbf{y}_i(t)=\sum_k \delta(t-t_{ik})$, with $t_{ik}$ being its $k^{\mathrm{th}}$ spike time.  Following \cite{kempter_hebbian_1999} we relate the joint statistics of $\mathbf{y}_i(t)$ and $\mathbf{y}_j(t)$ to the evolution of synaptic weights.  We first assume that individual pre-post spike pairs induce small changes in synaptic weights ($f_\pm \ll W^\mathrm{max}$).  This makes synaptic weights evolve slowly, on a much longer timescale than the millisecond scale of pairwise spiking covariance due to network interactions.  The separation of timescales between synaptic plasticity and spiking activity provides an approximation to the evolution of the synaptic weights (Methods: learning dynamics):
\begin{equation} \label{dWdt}
\frac{d\mathbf{W}_{ij}}{dt} = \mathbf{W}^0_{ij}\int_{-\infty}^\infty L(s)\big(r_ir_j + \mathbf{C}_{ij}(s)\big)ds.
\end{equation}
Here $r_i = \langle \mathbf{y}_i(t) \rangle$ is the time-averaged firing rate of neuron $i$, and $\mathbf{C}_{ij}(s) = \langle (\mathbf{y}_{i}(t)-r_i)(\mathbf{y}_{j}(t+s)-r_j) \rangle$ is the cross-covariance function of neuron $i$ and $j$'s spike trains.  The separation of timescales allows us to calculate the equilibrium spiking statistics $\mathbf{C}$, taking $\mathbf{W}$ to be constant on the timescale of $\mathbf{C}(s)$.  The term $r_ir_j$ in Eq. \eqref{dWdt} captures the firing rate dependence of STDP, while $\mathbf{C}_{ij}(s)$ models the sensitivity of STDP to spike timing.  Finally, $\mathbf{W}^0$ is the adjacency matrix of the network -- a binary matrix with $\mathbf{W}^0_{ij}=1$ denoting the presence of a synapse.  Multiplying by $\mathbf{W}^0_{ij}$ ensures that synapses that do not exist cannot potentiate into existence.  Eq. \eqref{dWdt} requires only the first and second order joint spiking statistics.  To facilitate calculations, many previous studies have used Poisson neuron models with a specified $r_i$ and $\mathbf{C}_{ij}(s)$ to generate $\mathbf{y}_i(t)$.  In contrast, we will use a white noise driven exponential integrate-and-fire model \cite{fourcaud-trocme_how_2003} for the generation of spike times (Methods: Network model).  While this complicates the calculation of the spike train statistics, it provides a more biophysically realistic model of neural dynamics \cite{jolivet_generalized_2004, jolivet_quantitative_2008} that better captures the timescales and neuronal nonlinearities that shape $r_i$ and $\mathbf{C}_{ij}(s)$.  In total, the above theory determines synaptic evolution from the integrated combination of an STDP rule $L(s)$ and the spike train cross-covariance function $\mathbf{C}_{ij}(s)$.  Thus, any mechanism affecting  two neurons' spiking covariance is expected to shape network structure through STDP.  

 %Furthermore, when correlations are weak there is a growing literature that provides opens a door to future examination of how biophysical properties, such as the variance or mean of synaptic inputs, the timescale of synaptic currents, or the presence of slow membrane conductances-- all of which shape spiking covariance -- interact with STDP to shape network structure \cite{shea-brown_correlation_2008, litwin-kumar_balanced_2011, ocker_Kv7_2014}.  

\paragraph{} 
 As a simple illustration of how spiking correlations can drive STDP, we examined the synaptic weight dynamics, $\mathbf{W}_{12}(t)$ and $\mathbf{W}_{21}(t)$, in a reciprocally coupled pair of neurons, both in the presence and absence of common inputs.  Specifically, the fluctuating input to neuron $i$ was the sum of a private and common term, $\sqrt{1-c}\xi_i(t)+ \sqrt{c}\xi_c(t)$, with $c$ the fraction of shared input to the neurons.  In the absence of common input ($c=0$; Fig. 1B), the two synapses behaved as expected with Hebbian STDP: one synapse potentiated and the other depressed (Fig. 1C).  In contrast, the presence of common input ($c=0.05$) was a source of synchrony in the two neurons' spike trains, inducing a central peak in the spike train cross-covariance function $\mathbf{C}_{ij}(s)$ (Fig. 1B vs 1D).  In this case, both synapses potentiated (Fig. 1E) because the common input increased synchronous spiking.  Since the potentiation side of the learning rule was sharper than the depression side (Fig. 1A), this enhanced the degree of overlap between $\mathbf{C}_{ij}(s)$ and the potentiation component of $L(s)$.  This overcame the effects of depression in the initially weaker synapse and promoted strong, bidirectional connectivity in the two-neuron circuit.

\paragraph{}
This example highlights how the temporal shape of the spike train cross-covariance function can interact with the shape of the learning rule, $L(s)$, to direct spike timing-dependent plasticity.  However, this case only considered the effect of correlated inputs from outside of the modeled circuit (Fig. 1).  %In the absence of correlated external inputs, the recurrent structure of neuronal networks is the only source of spiking covariance.  
Our primary goal is to predict how spiking covariance due to internal network interactions combines with STDP to drive self-organized network structure.  In order to do this, we first require a theory for predicting the spiking covariance between all neuron pairs given some static, recurrent connectivity.  Once this theory has been developed, we will use it to study the case of plastic connectivity.

\subsection*{Network architecture determines spiking covariance in static networks}  
\paragraph{}
In this section we review approximation methods \cite{doiron_oscillatory_2004, lindner_theory_2005, trousdale_impact_2012} that estimate the pairwise spike train cross-covariances $\mathbf{C}_{ij}(s)$ using a static weight matrix $\mathbf{W}$ (see Methods: Spiking statistics for a full description).  The exposition is simplified if we consider the Fourier transform of a spike train, $\mathbf{y}_i(\omega) = \int_{-\infty}^\infty \mathbf{y}_i(t)e^{-2\pi i \omega t}dt $, where $\omega$ is frequency.  Assuming weak synaptic connections $\mathbf{W}_{ij}$, we approximate the spike response from neuron $i$ as:
\begin{equation} \label{linear}
\mathbf{y}_i(\omega) =  \mathbf{y}_{i}^0(\omega) +\mathbf{A}_i(\omega)\left(\sum_{j=1}^N \mathbf{W}_{ij}J(\omega)\mathbf{y}_j(\omega) \right).
\end{equation} 	
The function $\mathbf{A}_i(\omega)$ is the linear response \cite{gardiner_stochastic_2009} of the postsynaptic neuron, measuring how strongly modulations in synaptic currents at frequency $\omega$ are transferred into modulations of instantaneous firing rate about a background state $ \mathbf{y}_{i}^0 $.   The function $J(\omega)$ is a synaptic filter.  In brief, Eq. \eqref{linear} is a linear ansatz for how a neuron integrates and transforms a realization of synaptic input into a spike train.         

\paragraph{}
Following \cite{doiron_oscillatory_2004, lindner_theory_2005, trousdale_impact_2012} we use this linear approximation to estimate the Fourier transform of $\mathbf{C}_{ij}(s)$, written as $\mathbf{C}_{ij}(\omega) = \langle \mathbf{y}_i(\omega) \mathbf{y}_{j}^*(\omega) \rangle$; here $\mathbf{y}^*$ denotes complex conjugation.  This yields the following matrix equation: 
\begin{equation} \label{Cfull}
\mathbf{C}(\omega) = \Big(\mathbf{I} -\big(\mathbf{W\cdot K}(\omega)\big)\Big)^{-1}\mathbf{C}^0(\omega) \Big(\mathbf{I}- \big(\mathbf{W\cdot K^*}(\omega)\big)\Big)^{-1},
\end{equation} 	
where $\mathbf{K}(\omega)$ is an interaction matrix defined by $\mathbf{K}_{ij}(\omega) = \mathbf{A}_i(\omega)\mathbf{J}_{ij}(\omega)$.  The matrix $\mathbf{C}^0(\omega)$ is the baseline covariance, with elements $\mathbf{C}_{ij}^0(\omega) = \langle \mathbf{y}_{i}^0(\omega) \mathbf{y}_{j}^{0*}(\omega) \rangle$, and $\mathbf{I}$ is the identity matrix.  Using Eq. \eqref{Cfull} we recover the matrix of spike train cross-covariance functions $\mathbf{C}(s)$ by inverse Fourier transformation.  Thus, Eq. \eqref{Cfull} provides an estimate of the statistics of pairwise spiking activity in the full network, taking into account the network structure. 

\paragraph{}
As a demonstration of the theory, we examined the spiking covariances of three neurons from a 1,000-neuron network (Fig. 2A, colored neurons).  The synaptic weight matrix $\mathbf{W}$ was static and had an adjacency matrix $\mathbf{W}^0$ that was randomly generated with Erd\"{o}s-R\'{e}nyi statistics (connection probability of 0.15).   The neurons received no correlated input from outside the network, making $\mathbf{C}^0(\omega)$ a diagonal matrix, and thus recurrent network interactions were the only source of spiking covariance.  %Nevertheless, the structure of network connections dictates the shape of spiking covariances.  
Neuron pairs that connected reciprocally with equal synaptic weights had temporally symmetric spike train cross-covariance functions (Fig. 2C), while uni-directional connections gave rise to temporally asymmetric cross-covariances (Fig. 2D).  When neurons were not directly connected, their covariance was weaker than that of directly connected neurons but was still nonzero (Fig. 2E).  The theoretical estimate provided by Eq. \eqref{Cfull} was in good agreement with estimates from direct simulations of the network (Fig. 2C,D,E red vs. gray curves).   

\subsection*{Self-consistent theory for network structure and spiking covariance with plastic synapses}  
\paragraph{}
In general, it is challenging to develop theoretical techniques for stochastic systems with several variables and nonlinear coupling \cite{gardiner_stochastic_2009}, such as in Eq. \eqref{dWdt}. Fortunately, in our model the timescale of spiking covariance in the recurrent network with static synapses is on the order of milliseconds (Fig. 2C,D,E), while the timescale of plasticity is minutes (Fig. 1C,E).   This separation of timescales provides an opportunity for a self-consistent theory for the coevolution of $\mathbf{C}(s)$ and $\mathbf{W}(t)$.  That is, so long as $f_{\pm}$ in Eq. \eqref{stdprule} are sufficiently small, we can approximate $\mathbf{W}$ as static over the timescales of $\mathbf{C}(s)$ and insert Eq. \eqref{Cfull} into Eq. \eqref{dWdt}.  The resulting system yields a solution $\mathbf{W}(t)$ that captures the long timescale dynamics of the plastic network structure (Methods: self-consistent theory for network plasticity).

\paragraph{}
As a first illustration of our theory, we focus on the evolution of three synaptic weights in a 1,000-neuron network (Fig. 3A, colored arrows).  The combination of Eqs. \eqref{dWdt} and \eqref{Cfull} predicted the dynamics of $\mathbf{W}(t)$, whether the weight increased with time (Fig. 3B left, red curve), decreased with time (Fig. 3C left, red curve), or remained approximately constant (Fig. 3D left, red curve).  In all three cases, the theory matched well the average evolution of the synaptic weight estimated from direct simulations of the spiking network (Fig. 3B,C,D left, thick black curves).  Snapshots of the network at three time points (axis arrows in Fig. 3B,C,D, left), showed that $\mathbf{W}$ coevolved with the spiking covariance (Fig. 3B,C,D right).  We remark that for any realization of background input $\mathbf{y}^0(t)$, the synaptic weights $\mathbf{W}(t)$ deviated from their average value with increasing spread (Fig. 3B,C,D left, thin black curves).  This is expected since $\mathbf{C}(t)$ is an average over realizations of $\mathbf{y}^0(t)$, and thus provides only a prediction for the drift of $\mathbf{W}(t)$, while the stochastic nature of spike times leads to diffusion of $\mathbf{W}(t)$ around this drift \cite{kempter_hebbian_1999}.  

\paragraph{}
In sum, the fast-slow decomposition of spiking covariance and synaptic plasticity provides a coherent theoretical framework to investigate the formation of network structure through STDP.  Also, our treatment is complementary to past studies on STDP \cite{gilson_emergence_2009-1, gilson_emergence_2009-2, gilson_emergence_2009-4} that focused on the development of architecture through external input.  In our model, there is no spiking covariance in the background state (i.e. $\mathbf{C}_{ij}^0(s)=\langle \mathbf{y}_i^0(t+s) \mathbf{y}_{j}^0(t) \rangle = 0 $ for $i \ne j$). Hence, we are specifically considering self-organization of network structure through internally generated spiking covariance.  

\paragraph{}
While our theory gives an accurate description of plasticity in the network, it is nevertheless high-dimensional.  Keeping track of every individual synaptic weight and spike train cross-covariance function involves $\mathcal{O}(N^2)$ variables.  For large networks, this becomes computationally challenging.  More importantly, this high-dimensional theory does not provide insights into the plasticity of the \emph{connectivity patterns} or \emph{motifs} that are observed in cortical networks \cite{song_highly_2005, perin_synaptic_2011}.  Motifs involving two or more neurons represent correlations in the network's weight matrix, which cannot be described by a straightforward  application of mean-field techniques.  In the next sections, we develop a principled approximation of the high-dimensional theory to a closed low-dimensional mean-field theory for how the mean weight and the strength of two-synapse motifs evolve due to STDP.  

\subsection*{Dynamics of mean synaptic weight}
\paragraph{}
We begin by considering the simple case of a network with independent weights (an Erd\"{o}s-R\'{e}nyi network).  In this case, the only dynamical variable characterizing network structure is the mean synaptic weight, $p$:
\begin{equation} \label{pdef}
p=\frac{1}{N^2}\sum_{i,j}\mathbf{W}_{ij}.
\end{equation}
The mean synaptic weight, $p$, in addition to the connection probability $p_0$, completely characterizes the connectivity of a weighted Erd\"{o}s-R\'{e}nyi network.  In order to calculate the dynamics of $p$, we insert the fast-slow STDP theory of Eq. \eqref{dWdt} into Eq. \eqref{pdef}:
\begin{equation}
\frac{dp}{dt} = \frac{1}{N^2}\sum_{i,j}\mathbf{W}^0_{ij} \int_{-\infty}^\infty L(s)\big(r_ir_j + \mathbf{C}_{ij}(s)\big)ds,
\end{equation}
where the spiking covariances are calculated using linear response theory (Eq. \eqref{Cfull}).  This equation depends on the network structure in two ways.  First, it depends on the full adjacency matrix $\mathbf{W}^0$.  Second, the spike train cross-covariances depend on the full weight matrix: $\mathbf{C}_{ij}(s) = \mathbf{C}_{ij}(s;\mathbf{W})$.  This dependence of a first--order connectivity statistic on the network structure poses a challenge for the development of a closed theory.

\paragraph{}
The main steps in our approach here are two approximations.  First, the matrix of spike train cross-covariances $\mathbf{C}(s)$ obtained from our linear ansatz (Eq. \eqref{Cfull}) can be expanded in a power series around the background cross-covariances $\mathbf{C}^0(s)$ (see Eq. \eqref{Cseries}).  Powers of the interaction matrix $\mathbf{K}$ in this series correspond to different lengths of paths through the network \cite{pernice_how_2011, trousdale_impact_2012}.    We truncate the spiking covariances at length one paths to obtain:
\begin{equation}
\mathbf{C}_{ij}(s) \approx \underbrace{\left(\mathbf{W}_{ij}\mathbf{K}_{ij}*\mathbf{C}^0_{jj}\right)(s)}_{\mathrm{forward}} + \underbrace{\left(\mathbf{C}^0_{ii}*\mathbf{W}_{ji}\mathbf{K}_{ji}^-\right)(s)}_{\mathrm{backward}} + \underbrace{\sum_{k}\left(\mathbf{W}_{ik}\mathbf{K}_{ik}*\mathbf{C}^0_{kk}*\mathbf{W}_{jk}\mathbf{K}_{jk}^- \right)(s)}_{\mathrm{common}}, \label{Ctrunc}
\end{equation}
where $*$ denotes convolution.  This truncation separates the sources of covariance between the spiking of neurons $i$ and $j$ into direct forward ($i \leftarrow j$) and backward ($i \rightarrow j$) connections, and common ($k \to i$ and $k \to j$) inputs.  %Neglecting contributions to $\mathbf{C}(s)$ from higher-order connection motifs removes the dependence of the two-synapse motifs on the full network structure through the spiking covariance.  
Nevertheless, after truncating $\mathbf{C}(s)$, the mean synaptic weight still depends on higher-order connectivity motifs (Eq. \eqref{dpdt2}).  Fortunately, for an Erd\"{o}s-R\'{e}nyi network, these higher-order motifs are negligible.  

\paragraph{}
The second approximation is to ignore the bounds on the synaptic weight in Eq. \eqref{stdprule}.   While this results in a theory that only captures the transient dynamics of $\mathbf{W}(t)$, it greatly simplifies the derivation of the low-dimensional dynamics of motifs, because dynamics along the boundary surface are not considered.  

\paragraph{}
With these two approximations, the mean synaptic weight obeys:
\begin{equation} \label{dpdt_sum}
\frac{dp}{dt} = r^2S \frac{1}{N^2}\sum_{i,j}\mathbf{W}^0_{ij} + S_F\frac{1}{N^2}\sum_{i,j}\mathbf{W}^0_{ij}\mathbf{W}_{ij} + S_B \frac{1}{N^2}\sum_{i,j}\mathbf{W}^0_{ij}\mathbf{W}_{ji} + S_C \frac{1}{N^2}\sum_{i,j,k} \mathbf{W}^0_{ij}\mathbf{W}_{ik}\mathbf{W}_{jk}.
\end{equation}
\paragraph{}
The first term on the right hand side of Eq. \eqref{dpdt_sum} is scaled by $S=\int_{-\infty}^\infty L(s)ds$, modeling the interaction between STDP rules that lead to net potentiation ($S>0$) or depression ($S <0$), and the mean firing rate, $r$, across the network.  This captures STDP due to chance spiking coincidence.  The remaining terms capture how synaptic weights interact with the temporal structure of spiking covariance. Because of the expansion in Eq. \eqref{Ctrunc}, these dependencies decompose into three terms, each scaled by the integral of the product of the STDP rule $L(s)$ and a component of the spike train cross-covariance $\mathbf{C}(s)$.  Specifically, covariance due to forward connections is represented by $S_F$ (Eq. \eqref{SFdef}; Fig. 4A), covariance due to backward connections is represented by $S_B$ (Eq. \eqref{SBdef}; Fig. 4B), and finally covariance due to common connections is represented by $S_C$ (Eq. \eqref{SCdef}; Fig. 4C).

\paragraph{}
For an Erd\"{o}s-R\'{e}nyi network, each sum in Eq. \eqref{dpdt_sum} can be simplified.  Let $p_0 = \frac{1}{N^2}\sum_{i,j}\mathbf{W}^0_{ij}$ be the connection probability of the network.  Since our theory for spiking covariances required weak synapses, we also explicitly scaled the weights, motifs, and amplitude of synaptic changes $f_\pm$ by $\epsilon = 1/(Np_0)$.  This ensured that as the connection probability $p_0$ was varied, synaptic weights scaled to keep the total input to a neuron constant (neglecting plasticity).  The first and second terms of Eq. \eqref{dpdt_sum} correspond to the definitions of $p_0$ and $p$.  Since different elements of $\mathbf{W}^0$ and $\mathbf{W}$ are independent, the third term reduces to $\frac{1}{N^2}\sum_{i,j}\mathbf{W}^0_{ij}\mathbf{W}_{ji} = \epsilon pp_0+ \mathcal{O}(\epsilon^{3/2})$ due to the central limit theorem.  The last term can be similarly evaluated and the dynamics of $p$ to first order in $\epsilon$ reduce to:
\begin{equation} \label{dpdt_1d}
\frac{dp}{dt} = p_0r^2 S +  \epsilon \left(p\left(S_F +  p_0S_B\right) + p^2S_C\right). %+\mathcal{O}(\epsilon^{3/2})
\end{equation}
 We next study this mean-field theory in two regimes, before examining the plasticity of non-Erd\"{o}s-R\'{e}nyi networks that exhibit motifs.

%The dynamics of $p(t)$ will involve the dynamics of different motifs through all of these components.    

\subsection*{Unbalanced STDP of the mean synaptic weight}
\paragraph{}
Eq. \eqref{dpdt_1d} contains one term proportional to product of firing rates and the integral of the STDP rule, $r^2 S$, and additional terms proportional to the small parameter $\epsilon$.  When the learning rule, $L(s)$, is dominated by either depression or potentiation (so that $S\sim \mathcal{O}(1)\gg\epsilon$) the whole network uniformly depresses (Fig. 5A,C) or potentiates (Fig. 5B,D) due to chance spike coincidences (the firing rate term dominates in Eq. \eqref{dWdt}).  These dynamics are straightforward at the level of individual synapses and this intuition carries over straightforwardly to the mean synaptic weight.  When the STDP rule is dominated by potentiation or depression, the $\mathcal{O}(\epsilon)$ terms in Eq. \eqref{dpdt_1d} are negligible; the average plasticity is solely determined by the firing rates, with spiking covariance playing no role.  In this case, the leading-order dynamics of $p$ are:
\begin{equation} \label{p_unbal}
p(t) = p_0r^2St + p(0),
\end{equation}
so that the mean synaptic weight either potentiates to its upper bound $p_0W^\mathrm{max}$ or depresses to $0$, depending on whether the integral of the STDP rule, $S$, is positive or negative.  For both depression and potentiation dominated STDP our simple theory in Eq. \eqref{p_unbal} quantitatively matches $p(t)$ estimated from simulations of the entire network (Fig. 5C,D, black vs. red curves). 

\subsection*{Balanced STDP of the mean synaptic weight}
\paragraph{}
On the other hand, if there is a balance between potentiation and depression in the STDP rule $L(s)$, then spiking covariance affects the average plasticity.  In order to make explicit the balance between potentiation and depression, we write $S = \pm \delta \epsilon$ (with $+\delta \epsilon$ for STDP with the balance tilted in favor of potentiation and $-\delta \epsilon$ for balance tilted in favor of depression).  %We use $\delta = .0001$.  
The leading-order dynamics of $p$ are then, for Erd\"{o}s-R\'{e}nyi networks,
\begin{equation} \label{dpdt_delepsdep}
\frac{1}{\epsilon}\frac{dp}{dt} = \pm\delta p_0r^2 +  p\left(S_F +  p_0S_B\right) + p^2S_C.
\end{equation}
%This quadratic equation admits two fixed points, one stable and the other unstable:
%\begin{equation}
%p = \frac{-\left(S_F+p_0S_B \right)\pm\sqrt{\left(S_F+p_0S_B \right)^2-4S_C\left(\pm\delta p_0r^2\right)}}{2S_C}
%\end{equation}
This quadratic equation admits up to two fixed points for $p$. We begin by examining the dynamics of $p$ for the case perfectly balanced potentiation and depression ($\delta = 0$) and a realistic shape of the STDP curve, and then consider the case of $\delta \neq 0$.  We find that $p$ potentiates for all cases except depression-dominated, balanced STDP, which can exhibit novel multistable dynamics.% The fixed points of $p$ are then at $0$ and $-\left(S_F+p_0S_B\right)/S_C$.  So the balance between $S_F$ and $p_0S_B$ and the sign of $S_C$ determine whether the second fixed point will be below or above zero, and whether it is attracting or repelling.  This depends on the shapes of the STDP rule and the spiking train cross-covariance functions.

\paragraph{}
Experimentally measured STDP rules in cortex often show $f_+ > f_-$ and $\tau_+ < \tau_-$ \cite{feldman_spike-timing_2012, froemke_spike-timing-dependent_2002}, making potentiation windows sharper and higher-amplitude that depression windows.  In this case, the STDP-weighted covariance from forward connections, $S_F>0$, is greater in magnitude than those from backward connections, $S_B<0$ (Fig. 4), and hence $S_F+p_0S_B > 0$.   Furthermore, since the covariance from common input decays symmetrically around $s=0$ (Fig. 4C), we have that $S_C >0$.  Consequently, when $\delta = 0$, all terms in Eq. \eqref{dpdt_delepsdep} are positive and $p$ potentiates to $p_0W^\mathrm{max}$. 

\paragraph{}
For potentation-dominated balanced STDP, $+\delta \epsilon$, again all terms in Eq. \eqref{dpdt_delepsdep} are positive and $p$ potentiates to $p_0W^\mathrm{max}$ (Fig. 6A).  However, with depression-dominated balanced STDP ($-\delta \epsilon$ in Eq. \eqref{dpdt_delepsdep}), $p$ has two fixed points, at:
\begin{equation} \label{p_delepsdep}
p = \frac{-\left(S_F+p_0S_B \right)\pm\sqrt{\left(S_F+p_0S_B \right)^2+4\delta p_0r^2S_C}}{2S_C}.
\end{equation}
Since $\left(S_F+p_0S_B\right)>0$ and $S_C>0$ because of our assumptions on $f_{\pm}$ and $\tau_{\pm}$, the term inside the square root is positive, and one fixed point is positive and the other negative.  The positive fixed point is unstable and, if within $[0,p_0W^\mathrm{max}]$, provides a separatrix between potentation and depression of $p$ (Fig. 6B).  This separatrix arises from the competition between potentiation (due to forward connections and common input) and depression (due to backward connections and firing rates).  
Examination of Eq. \eqref{p_delepsdep} shows competing effects of increasing $p_0$: it moves both fixed points closer to 0 (the $p_0S_B$ terms outside and inside the square root), but also pushes the fixed points away from 0 due to firing rate- and common-input terms.  The latter effect dominates for the positive fixed point.  So the mean synaptic weight of sparsely connected networks have a propensity to potentiate, while more densely connected networks are more likely to depress (Fig. 6B).

\paragraph{}
In total, we see that a slight propensity for depression can impose bistability on the mean synaptic weight.  In this case, a network with an initially strong mean synaptic weight $p(0)$ can overcome depression and strengthen synaptic wiring, while a network with the same STDP rule and connection probability but with an initially weak mean synaptic weight will exhibit depression.  In the next section we will show that similar separatrices exist in non-Erd\"{o}s-R\'{e}nyi networks with slightly depression-dominated STDP and govern motifs in such networks as well.

\section*{Motif dynamics}
\paragraph{}
We now consider networks that have structure at the level of motifs, so that different patterns of connectivity may be over- or under-represented compared to Erd\"{o}s-R\'{e}nyi networks.  We begin by defining the two-synapse motif variables: 
\begin{equation}
\begin{aligned} \label{motifdef}
q^\mathrm{div}&=\frac{1}{N^3}\sum_{i,j,k}\mathbf{W}_{ik}\mathbf{W}_{jk}-p^2,\\ 
q^\mathrm{con}&=\frac{1}{N^3}\sum_{i,j,k}\mathbf{W}_{ik}\mathbf{W}_{ij}-p^2, \\
q^\mathrm{ch}&=\frac{1}{N^3}\sum_{i,j,k}\mathbf{W}_{ij}\mathbf{W}_{jk}-p^2.
\end{aligned}
\end{equation}
The variables $q^\mathrm{div}$, $q^\mathrm{con}$ and $q^\mathrm{ch}$, respectively, measure the strength of divergent, convergent, and chain motifs.  For each variable, we subtract the expected value of the sum in a network with independent weights, $p^2$, so that the $q$s measure above- or below-chance levels of structure in the network.  Since these variables depend on the strength of both synapses making up the motif, we will refer to them as \emph{motif strengths}.   Motif strengths are also related to neurons' (weighted) in- and out-degrees (the total strength of incoming or outgoing synapses for each neuron).  $q^\mathrm{div}$ and $q^\mathrm{con}$ are proportional to the variance of neurons' in- and out-degrees.  $q^\mathrm{ch}$, on the other hand, is proportional to the covariance of neurons' in- and out-degrees.  This can be seen by taking the definitions of these motifs, Eq. \eqref{motifdef}, and first summing over the indices $i,j$.  This puts the sum in $q^\mathrm{div}$, for example, in the form of a sum over neurons' squared out-degrees.)

\paragraph{}
We now wish to examine the joint dynamics of the mean synaptic weight $p$ and these motif strengths.  In order to calculate the dynamics of, for example, $p$, we insert the fast-slow STDP theory of Eq. \eqref{dWdt} into the definition of $p$, as earlier.  Similarly to Eq. \eqref{dpdt_1d}, the dynamics of motifs $q^\mathrm{div}(t)$, $q^\mathrm{con}(t)$, and $q^\mathrm{ch}(t)$ then depend on the network structure.  This dependence of first- and second-order connectivity statistics on the network structure poses a challenge for the development of a closed theory for the dynamics of motifs.  The main steps in developing such a theory are the two approximations we used to develop Eq. \eqref{dpdt_1d}, as well as one more.

\paragraph{}
As above, our first approximation is to truncate the spike-train covariances at length one paths through the network.  This removes the dependency of the dynamics on longer paths through the network.  Nevertheless, after truncating $\mathbf{C}(s)$, the first- ($p$) and second-order ($q^\mathrm{div}$, $q^\mathrm{con}$, $q^\mathrm{ch}$) motifs still depend on higher-order motifs (Eq. \eqref{dpdt_sum}).  This is because of coupling between lower and higher-order moments of the connectivity matrix $\mathbf{W}$ (see Eqs. \eqref{dpdt2}-\eqref{dqxdivdt2}) and presents a significant complication.  

\paragraph{}
In order to close the dynamics at one- and two-synapse motifs, our new approximation follows \cite{hu_motif_2013}, and we rewrite higher-order motifs as combinations of individual synapses and two-synapse motifs (see Eqs. \eqref{resum1}-\eqref{resum2}).  For the mean synaptic weight, for example, one third-order motif appears due to the common input term of the spike-train covariances (Eq. \eqref{dpdt_sum}). 
% It corresponds to the strength of a divergent motif, $k \to i$ and $k \to j$, conditioned on the presence of a synapse $j \to i$.  This conditioning arises from the presence of the adjacency matrix $\mathbf{W}^0$ in Eq. \eqref{dWdt}.  $p$ measures the mean strength of connections $j \to i$.  So, the common input from a third neuron, $k$, only contributes to the plasticity if the connection $j \to i$ exists.  
 We break up this three-synapse motif into all possible combinations of two-synapse motifs and individual connections, estimating its strength as:
\begin{equation} \label{resum1a}
\frac{1}{N^3}\sum_{i,j,k}\mathbf{W}^0_{ij}\mathbf{W}_{ik}\mathbf{W}_{jk} \approx \left(p_0\left(q^\mathrm{div}+p^2\right) + p\left(q_\mathrm{X}^\mathrm{con} + q_\mathrm{X}^\mathrm{ch,B}\right)\right).
\end{equation}
This corresponds to assuming that there are no third- or higher-order correlations in the weight matrix beyond those due to second-order correlations; three- and more-synapse motifs are represented only as much as would be expected given the two-synapse motif strengths.  This allows us to close the motif structure at two-synapse motifs.  However, two new motifs appear in Eq. \eqref{resum1a}, $q_\mathrm{X}^\mathrm{con}$ and $q_\mathrm{X}^\mathrm{ch,B}$.  The $_\mathrm{X}$ subscript denotes that these motifs are mixed between the weight and adjacency matrices, measuring the strength of individual connections, conditioned on their being part of a particular motif.  $q_\mathrm{X}^\mathrm{con}$ corresponds to the strength of connections conditioned on being part of a convergent motif and $q_\mathrm{X}^\mathrm{ch,B}$ to the strength of connections conditioned on the postsynaptic neuron making another synapse in a chain (Eq. \eqref{motifdef1}).  As in previous sections, the final approximation is to ignore the bounds on the synaptic weight in Eq. \eqref{stdprule}, so that our theory only captures the transient dynamics of $\mathbf{W}(t)$.
%While this forces a theory that only captures the transient dynamics of $\mathbf{W}(t)$, it greatly simplifies the derivation of a low-dimensional theory of motifs because dynamics along the boundary surface are not considered. 

\paragraph{}
These approximations allow us (see Eqs. \eqref{dpdt1}, \eqref{dpdt2}, and \eqref{dpdtfull}) to rewrite the dynamics of the mean synaptic weight $p$ as:
\begin{equation} \label{dpdtresults}
\frac{dp}{dt} = p_0 r^2S + \epsilon \left[pS_F + \left(q_\mathrm{X}^\mathrm{rec}+p_0p\right)S_B + \frac{1}{p_0}\left(p_0\left(q^\mathrm{div}+p^2\right)+p\left(q_\mathrm{X}^\mathrm{con}+q_\mathrm{X}^\mathrm{ch,B}\right)\right)S_C \right].
\end{equation}
%The mean firing rate is $r$ and $p_0 = \frac{1}{N^2}\sum_{i,j}\mathbf{W}^0_{ij}$ is the connection probability of the network.  Since our theory for spiking covariances required weak synapses, we also explicitly scaled the weights, motifs, and amplitude of synaptic changes $f_\pm$ in Eq. \eqref{stdprule}, by $\epsilon = 1/(Np_0)$.  This ensured that as the connection probability $p_0$ was varied, synaptic weights scaled to keep the total input to a neuron constant (neglecting plasticity).  

\paragraph{}
The parameters $S$, $S_F$, $S_B$ and $S_C$ are defined as above.  Note that we recover Eq. \eqref{dpdt_1d} when all $q$'s vanish (i.e an Erd\"{o}s-R\'{e}nyi network).  When the network contains additional motifs ($q \ne 0$), the dynamics of $p$ contain new terms.  In Eq. \eqref{dpdtresults}, the influence of forward connections through $S_F$ is again proportional to the mean synaptic weight $p$.  In contrast, the influence of backward connections $S_B$ must interact with the new variable $q_\mathrm{X}^\mathrm{rec}$, which measures the mean strength of connections conditioned on their being part of a reciprocal loop (i.e the strength of a backwards connection, conditioned on the existence of the forward one).  As described above (Eq. \eqref{resum1a}), the covariance from common input $S_C$ involves $p$, the divergent motif, $q^\mathrm{div}$, as well as terms conditioned on weights being part of a convergent motif, $q_\mathrm{X}^\mathrm{con}$, or on the postsynaptic neuron making another synapse in a chain, $q_\mathrm{X}^\mathrm{ch,B}$.  The definitions for the mixed motifs, the $q_{\mathrm{X}}$s, are given in Eqs. \eqref{motifdef1}.  In total, the dynamics of mean synaptic weight cannot be written as a single closed equation, but also requires knowledge of how the second order motifs evolve.  

\paragraph{}
Fortunately, using a similar approach, dynamical equations can be derived for each of the two-synapse motifs $q^\mathrm{div}$, $q^\mathrm{cov}$, and $q^\mathrm{ch}$ (Eqs. \eqref{dqdivdtfull}-\eqref{dqchdtfull}). However, to close the system we require dynamics for five mixed motifs,  $q_\mathrm{X}^\mathrm{con}$,  $q_\mathrm{X}^\mathrm{div}$,  $q_\mathrm{X}^\mathrm{rec}$,  $q_\mathrm{X}^\mathrm{ch,A}$, and $q_\mathrm{X}^\mathrm{ch,B}$ (Eqs. \eqref{dqxrecdtfull}-\eqref{dqxchBdtfull}).  In total, this yields an autonomous 9-dimensional system of nonlinear differential equations describing the population averaged plasticity of second-order network structure.    We have derived these equations in the absence of common external inputs to the neurons; the theory can easily be extended to this case by including external covariance in Eq. \eqref{Ctrunc} (replacing $\mathbf{C}^0$ with $\left(\mathbf{C}^0 + \mathbf{C}^\mathrm{ext}\right)$, where $\mathbf{C}^\mathrm{ext}$ is the covariance matrix of the inputs).

\paragraph{}
When the network structure, $\mathbf{W}^0$ is approximately Erd\"{o}s-R\'{e}nyi, the motif frequencies $q_0$ are $\mathcal{O}\left(N^{-3/2}\right)=\mathcal{O}\left(\epsilon^{3/2}\right)$.  If we further assume initial conditions for the motif strengths and the mixed motifs to be consistent with Erd\"{o}s-R\'{e}nyi statistics ($q(0)\sim\mathcal{O}\left(\epsilon^{3/2}\right)$ for all motifs), then we also have $dq_\mathrm{X}/dt \sim \mathcal{O}\left(\epsilon^{3/2}\right)$ and $dq_\mathrm{X}/dt \sim \mathcal{O}\left(\epsilon^{3/2}\right)$ for each motif, Eqs. \eqref{dqdivdtfull}-\eqref{dqxchBdtfull}.  In this case we can neglect, to leading order, the dynamics of motifs entirely.  Thus, the set of Erd\"{o}s-R\'{e}nyi networks $\{p(t), q^\mathrm{div} = q^\mathrm{con} = q^\mathrm{ch} =  q_\mathrm{X}^\mathrm{rec} = q_\mathrm{X}^\mathrm{con} = q_\mathrm{X}^\mathrm{div} = q_\mathrm{X}^\mathrm{ch,A} = q_\mathrm{X}^\mathrm{ch,B} = 0\}$ forms an invariant set under the dynamics of the motif strengths; we examined the behavior on this invariant set in the above section (Fig. 5 and 6).

\paragraph{}
We refer to the mean field theory of Eqs. \eqref{dpdtfull}-\eqref{dqxchBdtfull} as the \textit{motif dynamics} for a recurrent network with STDP.  This theory accurately predicts the transient dynamics of the first- and two-synapse motifs of the full stochastic spiking network (Fig. 7, compare red versus thin black curves), owing to significant drift compared to diffusion in the weight dynamics and these network-averaged motif strengths.  The derivation and successful application of this reduced theory to a large spiking network %the problem of self-organization of network structure in spiking networks with plastic synapses 
is a central result of our study.  

\paragraph{}
Our theory captures several nontrivial aspects of the evolution of network structure.  First, while the STDP rule is in the depression-dominated regime ($S<0$ for the simulations in Fig. 7), the mean synaptic weight $p$ nevertheless grows (Fig. 7A).  Second, both divergent and convergent connections, $q^\mathrm{div}$ and $q^\mathrm{con}$, grow above what is expected for a random (Erd\"{o}s-R\'{e}nyi) graph (Fig. 7B,C); however, at the expense of chain connections $q^\mathrm{ch}$ which decay (Fig. 7G).  In the subsequent sections, we leverage the simplicity of our reduced theory to gain insight into how the STDP rule $L(s)$ interacts with recurrent architecture to drive motif dynamics.

\subsection*{Unbalanced STDP of two-synapse motifs}
%\paragraph{}
%The motif dynamics each contain one term proportional to the firing rates and integral of the STDP rule, $r^2 S$, and additional terms proportional to the small parameter $\epsilon$.  When the learning rule, $L(s)$, is dominated by either depression or potentiation (so that $S\sim \mathcal{O}(1)\gg\epsilon$) the whole network uniformly depresses (Fig. 5A,C) or potentiates (Fig. 5B,D) due to chance spike coincidences (the firing rate term dominates in Eq. \eqref{dWdt}).  These dynamics are straightforward at the level of individual synapses, but what do they correspond to for the motifs? 

\paragraph{}
When the STDP rule is dominated by potentiation or depression so that $S\sim \mathcal{O}(1)\gg\epsilon$, then the $\mathcal{O}(\epsilon)$ terms in Eqs. \eqref{dqdivdtfull}-\eqref{dqxchBdtfull} are negligible.  In this case each motif's plasticity is solely determined by the firing rates, with spiking covariance plays no role.  Here the motif dynamics are simply:
\begin{equation} \begin{aligned}
\frac{dp}{dt} &= p_0 r^2S + \mathcal{O}(\epsilon) \\
\frac{dq^\alpha}{dt} &= 2r^2Sq_\mathrm{X}^\alpha + \mathcal{O}(\epsilon) \\
\frac{dq_\mathrm{X}^\alpha}{dt} &= r^2Sq_0^\alpha + \mathcal{O}(\epsilon)
\end{aligned} \end{equation}
for $\alpha = \mathrm{div, con, or ch}$ (and taking $q_\mathrm{X}^\mathrm{ch} = \left(q_\mathrm{X}^\mathrm{ch,A} + q_\mathrm{X}^\mathrm{ch,B}\right)/2$ in the second equation).  The dynamics of $p$ are the same here as for the Erd\"{o}s-R\'{e}nyi case above; we include it for completeness.  Dropping order $\epsilon$ terms gives the simple solutions:
\begin{equation} \begin{aligned}
p(t) &= p_0r^2S t + p(0) \\
q^\alpha (t) &= q^\alpha(0) + q_\mathrm{X}^\alpha(0)r^2St + \frac{1}{2}q_0^\alpha \left(r^2S\right)^2 t^2
\end{aligned} \end{equation}
for $\alpha = \mathrm{div, con, ch}$ (Methods: Unbalanced STDP).  As stated previously, with $S \sim \mathcal{O}(1)$, individual synapses uniformly potentiate or depress (Fig. 5).  This is reflected in the linear decay or growth (for depression- or potentiation-dominated $L(s)$, respectively) of $p$ with $r^2$ and quadratic amplification of baseline motif frequencies for the two-synapse motif strengths.
%This gives linear decay or growth (for depression- or potentiation-dominated $L(s)$, respectively) of $p$ with $r^2$, and the two-synapse motif strengths quadratically amplify the frequencies of these motifs in the baseline network structure $\mathbf{W}^0$.  This reflects the uniform plasticity of individual synapses.  
%So while individual weights are uniformly potentiating or depressing, the strength of motifs depends on their baseline frequency.  

%\begin{equation}
%q_\mathrm{X}(t) = r^2Sq_0 t + q_\mathrm{X}(0)
%\end{equation}
%so that 
%\begin{align} \label{unbal}
%p(t) &= p_0r^2S t + p(0) \\
%q^\mathrm{div}(t) &= q^\mathrm{div}(0) + q_\mathrm{X}^\mathrm{div}(0)r^2St + \frac{1}{2}q_0^\mathrm{div} \left(r^2S\right)^2 t^2 \\
%q^\mathrm{con}(t) &= q^\mathrm{con}(0) + q_\mathrm{X}^\mathrm{con}(0)r^2St +  \frac{1}{2}q_0^\mathrm{con} \left(r^2S\right)^2 t^2 \\
%q^\mathrm{ch}(t) &= q^\mathrm{ch}(0) + \left(q_\mathrm{X}^\mathrm{ch,A}(0) + q_\mathrm{X}^\mathrm{ch,B}(0)\right)r^2St + \frac{1}{2}q_0^\mathrm{ch} \left(r^2S\right)^2 t^2
%\end{align}

\subsection*{Balanced STDP of two-synapse motifs}
\paragraph{}
Now, we turn our attention to how internally generated spiking covariance interacts with balanced STDP to control motifs (examining the dynamics of Eqs. \eqref{dpdtfull}-\eqref{dqxchBdtfull}).  As before, we consider STDP rules with sharper windows for potentiation than depression ($\tau_+ < \tau_-$ and $f_+ > f_-$).  Each two-synapse motif can have a nullcline surface in the nine-dimensional motif space.  These nullclines define a separatrix for the promotion or suppression of the corresponding motif, analogous to the case on the Erd\"{o}s-R\'{e}nyi invariant set (Fig. 7).  We illustrate this by examining the dynamics in the $\left(q^\mathrm{div},q^\mathrm{con}\right)$ plane.  For STDP rules with a balance tilted towards depression ($-\delta \epsilon$), the nullclines provided thresholds for the promotion or suppression of divergent or convergent motifs (Fig. 8A, blue lines).  The flow in this slice of the motif space predicted the motif dynamics well (Fig. 8A, compare individual realizations of the full spiking network -- thin black lines -- to the flow defined by the vector field of the reduced motif system).

\paragraph{}
For STDP rules with the balance tilted towards potentiation ($+\delta \epsilon$), on the other hand, the nullclines were at negative motif strengths (Fig. 8B).  Can the motif strengths achieve negative values? As stated previously, $q^\mathrm{con}$ and $q^\mathrm{div}$ are proportional to the variances of neurons' in and out degrees, respectively.  So, like the mean synaptic weight, $q^\mathrm{div}, q^\mathrm{con} \geq 0$, and these motifs always potentiated for $+\delta\epsilon$ STDP rules (Fig. 8B).

%When the two-synapse motifs are accounted for, there is an unstable manifold in the 9-dimensional motif space, analogous to the unstable fixed point on the invariant set.  Since it is off of the origin, this unstable manifold defines a separatrix for the promotion or suppression of different motifs.  We illustrate this by examining a slice through motif space: the $\left(q^\mathrm{div},q^\mathrm{con}\right)$ plane.  For STDP rules with a balance tilted towards depression ($-\delta \epsilon$), the unstable manifold provides a threshold for the promotion or suppression of divergent or convergent motifs (Fig. 8A, intersection of red lines).  The flow in this slice of the motif space predicted the motif dynamics well (Fig. 8A, compare individual realizations of the full spiking network -- thin black lines -- to the flow defined by the vector field of the reduced motif system).  
%This was the case even though the vector field was calculated by assuming that all other seven variables in the system were held fixed at their initial conditions \alkcomment{Why is this important?  It seems like it means that the dynamics aren't very interesting -- not much coupling between different variables.}.

\paragraph{} 
Chain motifs, in contrast, correspond to the covariance of neurons' weighted in- and out-degrees and so can achieve negative values.  Indeed, the strength of chains can depress below zero even while the mean synaptic weight and other motifs potentiate (Fig. 5A,G).  Examining how $q^\mathrm{ch},q^\mathrm{div}$ and $q^\mathrm{con}$ coevolve allowed us to see how in- and out-hubs developed in the network.  With the $+\delta\epsilon$ STDP rule, $q^\mathrm{ch}$ increased along with $q^\mathrm{con}$ and $q^\mathrm{div}$ (Figs. 8B, 9C,D).  So, individual neurons tended to become both in- and out-hubs.  With the $-\delta\epsilon$ STDP rule, however, $q^\mathrm{ch}$ could decrease while $q^\mathrm{div}$ and $q^\mathrm{con}$ increased (Fig. 5).  In this case, neurons tended to become in- or out-hubs, but not both.

\paragraph{}
Many studies have examined how STDP affects either feedforward or recurrent structure in neuronal networks, commonly showing that STDP promotes feedforward structure at the expense of recurrent loops \cite{fiete_spike-time-dependent_2010,kozloski_theory_2010,kunkel_limits_2011}.   This is consistent with the intuition gained from isolated pairs of neurons, where STDP can induce competition between reciprocal synapses and eliminate disynaptic loops (Supp. Fig. 1; \cite{song_competitive_2000}).  Our theory provides a new way to examine how STDP regulates feedforward vs recurrent motifs by examining the dynamics of $q^\mathrm{ch}$.  This variable includes both recurrent loops ($q^\mathrm{rec}$) and feedforward chains ($q^\mathrm{ff}$).  In order to understand the contribution of each of these to overall potentiation or depression of chains, we split the motif strength for chains into contributions from recurrent loops and feedforward chains, rewriting $q^\mathrm{ch}$ as:
\begin{equation} \label{ff_rec}
q^\mathrm{ch} = \underbrace{\frac{1}{N^3}\sum_{i,j,k}\delta_{ik}\mathbf{W}_{ij}\mathbf{W}_{jk}}_{q^\mathrm{rec}}  + \underbrace{\frac{1}{N^3}\sum_{i,j,k}\left(1-\delta_{ik} \right)\mathbf{W}_{ij}\mathbf{W}_{jk} - p^2}_{q^\mathrm{ff}}.
\end{equation}
%where
%\begin{equation} \begin{aligned}
%q^\mathrm{rec} &= \frac{1}{N^3}\sum_{i,j,k}\delta_{ik}\mathbf{W}_{ij}\mathbf{W}_{jk} = \frac{1}{N^3}\sum_{i,j} \mathbf{W}_{ij}\mathbf{W}_{ji} \\
%q^\mathrm{ff} &= \frac{1}{N^3}\sum_{i,j,k}\left(1-\delta_{ik} \right)\mathbf{W}_{ij}\mathbf{W}_{jk} - p^2
%\end{aligned} \end{equation}
Similar to the case of other two-synapse motifs, the leading order dynamics of the recurrent motif are:
\begin{equation}
\frac{1}{2\epsilon}\frac{dq^\mathrm{rec}}{dt} = r^2 Sp_0\left(q_\mathrm{X}^\mathrm{rec} + pp_0\right)+S_Fq^\mathrm{rec}+S_	Bq_\mathrm{X2}^\mathrm{rec}.
\end{equation}
We obtain the dynamics of the feedforward motif by subtracting $dq^\mathrm{rec}/dt$ from $dq^\mathrm{ch}/dt$ (Eq. \eqref{dqffdt}).  In Eq. \eqref{ff_rec} we subtract $p^2$ from $q^\mathrm{ff}$ because $q^\mathrm{ff}$ is the dominant contributor to $q^\mathrm{ch}$.  This restricts $q^\mathrm{rec}$ to being non-negative.  The new auxiliary variable $q_\mathrm{X2}^\mathrm{rec}$ is proportional to the conditional second moment of weights that are part of loops (Eq. \eqref{qx2rec}), and evolves according to Eq. \eqref{dqX2recdt}.  The replacement of $q^\mathrm{ch}$ by these variables expands the motif space to 11 dimensions.  
%If $q^\mathrm{ch}$ were held constant, this decomposition would induce a competition between loops and feedforward motifs.  Since $q^\mathrm{ch}$ varies, however, there does not have to be competition between loops and feedforward structure, contradicting the intuition gained from pairs of neurons \cite{song_competitive_2000}.  

\paragraph{}
We investigated the joint dynamics of feedforward chains and recurrent loops similarly to the other motifs, examining the $(q^\mathrm{ff},q^\mathrm{rec})$ plane.  The $q^\mathrm{ff}$ and $q^\mathrm{rec}$ nullclines divided this plane into regions where each motif potentiated or depressed.  The shape of the STDP rule and the initial values of the other motif strengths determined the location of these nullclines.  For the $+\delta\epsilon$ STDP rule, the $q^\mathrm{rec}$ nullcline was just below $q^\mathrm{rec}=0$ (Fig. 9A, blue horizontal line).  Since $q^\mathrm{rec}\geq 0$, this forced $q^\mathrm{rec}$ to potentiate.  The feedforward motif, in contrast, could potentiate or depress above chance levels.  In our spiking simulations, the initial conditions put $q^\mathrm{ff}$ in the region of depression, so that feedforward structure depressed even while all other motifs were growing (Fig. 9A, right panels).  

\paragraph{}
These dynamics were the opposite of what would be expected from examining isolated pairs of neurons.  With both the $+\delta\epsilon$ and $-\delta\epsilon$ balanced STDP rules, isolated pairs of neurons showed splitting of synaptic weights to eliminate the recurrent loop (Supp. Fig. 1).  Thus, with the $+\delta\epsilon$ STDP rule, the intuition gained from pairs of neurons did not predict the plasticity of feedforward and recurrent motifs.

\paragraph{}
The locations of the $q^\mathrm{ff}$ and $q^\mathrm{rec}$ nullclines were sensitive to the values of the other motif variables.  Since the mean synaptic weight and $q^\mathrm{div}$ and $q^\mathrm{con}$ exhibited bistability under the $-\delta\epsilon$ STDP rule, we examined the $(q^\mathrm{ff},q^\mathrm{rec})$ slice through motif space when the other motifs were potentiating (Fig. 9B, right panels) or depressing (Fig. 9C, right panels).  In both cases, the $q^\mathrm{rec}$ nullcline was above $0$ so that the recurrent motif could either potentiate or depress, depending on its initial strength (Fig. 9B,C blue horizontal lines).  Similarly, the feedforward motif could either potentiate or depress.  

\paragraph{}
In spiking simulations with $-\delta\epsilon$ STDP where $p$ and the other motifs potentiated (Fig. 9B, right), the initial conditions put $(q^\mathrm{ff},q^\mathrm{rec})$ in the region of phase space where they both depressed (Fig. 9B, left).  In spiking simulations with $-\delta\epsilon$ STDP where $p$ and other motifs depressed (Fig. 9C, right), the initial conditions put $(q^\mathrm{ff},q^\mathrm{rec})$ in the region where $q^\mathrm{ff}$ potentiated and $q^\mathrm{rec}$ depressed.  This region corresponded to what would be expected from examining pairs of neurons (Supp. Fig. 1): the loss of disynaptic loops and promotion of feedforward structure.  So with the $-\delta \epsilon$ STDP rule, the region of phase space where the pair-based intuition was accurate at the network level was accessible.  In most of the motif space, however, interactions between triplets of neurons played a strong role so that the theory developed here was necessary to predict the STDP of motif structure.

\section*{Discussion}
\paragraph{}
We have developed a theory for spike timing-dependent plasticity in weakly coupled recurrent networks of exponential integrate-and-fire neurons.  We used this framework to derive a low-dimensional dynamical system capturing the plasticity of two-synapse motifs.  The resulting system naturally classifies STDP rules into two categories: 1) rules with an imbalance between potentiation and depression whose dynamics are dominated by the firing rates of neurons in the network, and 2) rules with balanced potentiation and depression in which different sources of spiking covariance interact with the STDP rule to determine network structure.  In the latter case, any mechanism controlling spiking covariance in the network may affect how the network structure evolves.  Thus, spike initiation dynamics \cite{galan_correlation-induced_2006, de_la_rocha_correlation_2007, shea-brown_correlation_2008, hong_single_2012}, spike-frequency adaptation \cite{ocker_kv7_2014, deger_dynamics_2013}, synaptic inhibition \cite{renart_asynchronous_2010, brunel_dynamics_2000, litwin-kumar_spatial_2012} and passive membrane properties \cite{litwin-kumar_balanced_2011} could all, in addition to controlling firing rates, drive motif dynamics.
	
\subsection*{STDP in recurrent networks}
\paragraph{}
A recent suite of studies derived a theory for how STDP shapes the full structure of networks of neurons whose spike trains are Poisson processes \cite{burkitt_spike-timing-dependent_2007, gilson_emergence_2009-1, gilson_emergence_2009-2, gilson_emergence_2009-3, gilson_emergence_2009-4, gilson_emergence_2010}.  The initial approach is similar to ours with a linear approximation to estimate spiking covariance (see Eq. \eqref{linear}-\eqref{Cfull}).  However, these studies mostly focused on the effects of external input, considering how feedforward inputs entrain structure in recurrent synapses \cite{gilson_emergence_2009-1, gilson_emergence_2009-2, gilson_emergence_2009-4}.  The only source of spiking covariance was inherited from external sources (meaning ${\bf C}_0(s)$ has off-diagonal structure), and subsequently filtered by the network to produce spiking covariance.   Two previous studies by the same authors also examined STDP in networks without external stimuli \cite{burkitt_spike-timing-dependent_2007, gilson_emergence_2009-3}; however, these took a large system size limit ($ N \to \infty$) and neglected the contribution of spiking covariance to STDP, focusing on the firing rate dependence due to an unbalanced learning rule.  

\paragraph{}
In contrast, we consider the case where the intrinsic variability of neurons' spike trains is the only source of spiking covariance, necessitating a finite sized network ($\epsilon = 1/(N p_0) >0$).  There is little difference between our results and those of past studies \cite{burkitt_spike-timing-dependent_2007, gilson_emergence_2009-3} when the learning rule is unbalanced.  However, if there is a balance between potentiation and depression, our theory shows how internally generated spiking covariances play a strong role in STDP.  Furthermore, our use of integrate-and-fire models allows our theory to predict the evolution of network structure without fixing the statistics of individual or joint spiking activity.  

\subsection*{Stability of learned network structures}
\paragraph{}
Early studies of long-term plasticity, which gave rise to the phenomenological plasticity model we used, focused on the relative timing of action potentials.  More recent experiments have shown that neurons' firing rates and the postsynaptic membrane voltage and spike patterns all affect the shape of measured STDP curves \cite{sjostrom_rate_2001, bi_temporal_2002, froemke_spike-timing-dependent_2002, wang_coactivation_2005, wittenberg_malleability_2006}.  More complicated models of long-term plasticity, based on spike-triplet- or voltage-dependent STDP \cite{pfister_triplets_2006, clopath_claudia_connectivity_2010} or on calcium thresholds for the induction of depression and potentiation \cite{shouval_unified_2002, rubin_calcium_2005, graupner_calcium-based_2012}, can replicate many of these complexities.  The observation that firing rates undergo large fluctuations over slow timescales \cite{churchland_variance_2011, kohn_stimulus_2005, churchland_stimulus_2010, arieli_dynamics_1996, tsodyks_linking_1999} suggests that \emph{in vivo} STDP may transition between unbalanced potentiation- and depression-dominated regimes.  While long-term plasticity can be strongly affected by pre- and postsynaptic firing rates, connectivity motifs and spiking covariance could determine the direction of plasticity during transitions between potentiation- and depression-dominated regimes.  While our paper provides an initial framework to study how STDP shapes structure in recurrent networks, a more realistic learning rule than that used here (Eq. \eqref{stdprule}) will be needed to address these issues.

\paragraph{}
The additive, Hebbian STDP model we used here gives rise to splitting of synaptic weights: individual weights potentiate to some upper bound, or depress to a lower bound.  This produces a bimodal distribution of synaptic weights, while experimentally observed weight distributions tend to be unimodal and long-tailed \cite{song_highly_2005, perin_synaptic_2011, lefort_excitatory_2009, ikegaya_interpyramid_2013}.  Modifications of this model, such as introducing axonal or dendritic delays or weight-dependence of plasticity, can yield weight distributions more closely resembling those observed in neural tissue \cite{rubin_equilibrium_2001, rubin_steady_2001, guetig_learning_2003, babadi_intrinsic_2010, gilson_stability_2011}.  Depending on the modification made (delays vs weight-dependence), either the same or similar theories for motif plasticity can be derived using the methods presented in our study.  Strong weight dependence, however, forces every weight to the same value so that the baseline motif frequencies completely determine the structure of the weight matrix (Supplemental Information: Multiplicative STDP).  The dynamics of motifs under more realistic models of synaptic plasticity remain to be studied.	

\paragraph{}
A major feature of STDP is that it can potentiate temporally correlated inputs \cite{kempter_hebbian_1999}.  Since synchronous inputs are effective at driving postsynaptic spiking, this can give rise to pathological activity in recurrent networks \cite{morrison_spike-timing-dependent_2007}.  Synaptic depression driven by postsynaptic spikes, independent of presynaptic activity, can stabilize postsynaptic firing rates during STDP \cite{kempter_intrinsic_2001, gilson_emergence_2009-1}.  Such additional rate-dependent terms of the plasticity rule can also stabilize the full weight matrix \cite{gilson_emergence_2009-3} and thus give rise to stable motif configurations.  Recent work has focused on the necessity of homeostatic mechanisms, including synaptic scaling \cite{royer_conservation_2003} or inhibitory plasticity, in stabilizing both the activity and structure of neural networks \cite{renart_robust_2003, lazar_sorn:_2009, zheng_network_2013, fiete_spike-time-dependent_2010, zenke_synaptic_2013, litwin-kumar_formation_2014}.  Since balanced STDP can give rise to bistability of mean synaptic weights in a network (Fig. 7B), it could also provide a mechanism for assembly formation (selected weights potentiate, while other weights depress).  Mechanisms of metaplasticity \cite{abraham_metaplasticity:_2008}, operating on a similar timescale to STDP, could give rise to such a balance.  This suggests a novel role for metaplasticity in controlling not only single-neuron excitability but also the self-organization of microcircuits in recurrent networks.  

\subsection*{Plasticity of motifs}
\paragraph{}
Early studies on STDP focused on isolated pairs of reciprocally connected neurons, showing that the type of STDP we study tends to induce competition between reciprocal synapses (Fig. 1B,C; \cite{song_competitive_2000}).
Since then, many simulation studies have investigated how STDP affects the structure and activity of recurrent networks \cite{izhikevich_spike-timing_2004, levy_distributed_2001, mongillo_learning_2005, liu_embedding_2009,litwin-kumar_formation_2014}, commonly examining the emergence of highly connected clusters.  Reduced theories exposing how STDP shapes network-level structure have, however, been difficult to obtain.  Most have examined the average synaptic weight in a network \cite{chen_mean-field_2010, mayer_dynamical_2012}, focusing on the relationship between network-averaged firing rates and mean synaptic weights ($p$) but neglecting spiking covariance.  Mean-field theories are accurate for fully homogenous networks; however, if all neurons have the same weighted in- and out-degrees, there is no plasticity of two-synapse motifs (Supplemental Information: Motif plasticity in homogenous networks).  So plasticity of higher-order network structure depends on inhomogeneities in neurons' inputs and outputs.

\paragraph{}
The few reduced theories examining STDP of higher-order network structure have focused on the question of how STDP controls feedforward chains versus recurrent loops.  One study compared the mean strengths of feedforward versus recurrent inputs in a network receiving synchronous stimulation \cite{kunkel_limits_2011}, but did so for a neuron that made no feedback connections to the network -- effectively only taking into account the first term of Eq. \eqref{Ctrunc}. Another study examined the strength of loops in a network of linear excitatory neurons, showing that STDP tends to reduce the total number of loops (of all lengths) in a network \cite{kozloski_theory_2010}.  Our theory is restricted to two-synapse loops; while we have shown that these can potentiate (as in Fig. 9C), \cite{kozloski_theory_2010} predicts that longer loops would meanwhile be weakened.  Whether this is the case with balanced STDP driven by more realistic neuron models remains to be seen.  

\paragraph{}
There is a growing body of evidence that cortical networks exhibit fine-scale structure \cite{markram_network_1997, perin_synaptic_2011, song_highly_2005, yoshimura_excitatory_2005}.  Experimental studies have shown that such microcircuits depend on sensory experience \cite{ko_emergence_2013, ko_emergence_2014}.  Our work provides an important advance towards explicitly linking the plasticity rules that control individual synapses and the emergent microcircuits of cortical networks.  We have shown that synaptic plasticity based only on temporally precise spike-train covariance can give rise to a diversity and, under certain conditions, multistability of motif configurations.  Motifs can have a strong influence on pairwise and population-level activity \cite{zhao_synchronization_2011, roxin_role_2011, litwin-kumar_slow_2012, gaiteri_interaction_2011, kriener_correlations_2009, pernice_how_2011, pernice_relevance_2013, trousdale_impact_2012, hu_motif_2013, helias_correlation_2014, hu_local_2014}, suggesting that precise spike timing may play a role in how networks reorganize patterns of connectivity in order to learn computations.

\section*{Methods}
\subsection*{Neuron and network model}
\paragraph{}
We model a network of $N$ neurons.  The membrane dynamics of individual neurons obey the exponential integrate-and-fire (EIF) model \cite{fourcaud-trocme_how_2003}, one of a class of models well-known to capture the spike initiation dynamics and statistics of cortical neurons \cite{jolivet_generalized_2004, jolivet_quantitative_2008}.  Specifically, the membrane voltage of neuron $i$ evolves according to:
\begin{equation} \label{dVdt}
C\frac{dV_i}{dt} = g_L\left(V_L-V_i \right) + g_L\Delta\exp{\left(\frac{V_i-V_T}{\Delta}\right)} + I_\mathrm{i}(t) + \sum_{j=1}^N \mathbf{W}_{ij}\left(\mathbf{J}_{ij}*y_j.\right).
\end{equation}
The first term on the right-hand side is the leak current, with conductance $g_L$ and reversal potential $V_L$.  The next term describes a phenomenological action potential with an initiation threshold $V_T$ and steepness $\Delta$: when the voltage reaches $V_T$, it diverges; this divergence marks an action potential.  For numerical simulations, action potentials are thresholded at $V(t) = V_\mathrm{th}$, reset to a reset potential $V_{re}$ and held there for an absolute refractory period $\tau_{\mathrm{ref}}$.

\paragraph{}
Input from external sources not included in the model network is contained in $I_\mathrm{i}(t)$.  We model this as a Gaussian white noise process: $I_i(t) = \mu + g_L\sigma D \xi_i(t)$.  The mean of the the external input current is $\mu$.  The parameter $\sigma$ controls the strength of the noise and $D = \sqrt{\frac{2C}{g_L}}$ scales the noise amplitude to be independent of the passive membrane time constant.  With this scaling, the infinitesimal variance of the passive membrane voltage is $\left(g_L\sigma D\right)^2$.

\paragraph{}
The last term of Eq. \eqref{dVdt} models synaptic interactions in the network.  The $N\times N$ matrix $\mathbf{W}$ contains the amplitudes of each synapse's postsynaptic currents.  It is a weighted version of the binary adjacency matrix $\mathbf{W}^0$, where $\mathbf{W}^0_{ij} = 1 (0)$ indicates the presence (absence) of a synapse from neuron $j$ onto neuron $i$.  If a synapse $ij$ is present then $\mathbf{W}_{ij}$ denotes its strength.  Due to synaptic plasticity, $\mathbf{W}$ is dynamic; it changes in time as individual synapses potentiate or depress.  The spike train from neuron $j$ is the point process $y_j(t) = \sum_k \delta(t-t_j^k)$, where $t_j^k$ denotes the $k^{\mathrm{th}}$ spike time from neuron $j$.  The $N\times N$ matrix $\mathbf{J}(t)$ defines the shape of the postsynaptic currents.  In this study, we use exponential synapses: $\mathbf{J}_{ij}(t-t_j^k) = \mathcal{H}(t-t_j^k)\exp{\left(-\frac{t-t_j^k}{\tau_\mathrm{S}}\right)}$, where $\mathcal{H}(t)$ is the Heaviside step function.  Our theory is not exclusive to the EIF model or to the simple synaptic kernels we used; similar methods can be used with any integrate-and-fire model and arbitrary synaptic kernels.  Model parameters are contained in Table 1 (unless specified otherwise in the text).  

Unless otherwise stated we take the adjacency matrix ${\bf W}_0$ to have Erd\"{o}s-R\'{e}nyi statistics with connection probability $p_0=0.15$. 
 
\subsection*{Learning dynamics}
\paragraph{}
We now derive Eq. \eqref{dWdt}, summarizing a key result of \cite{kempter_hebbian_1999}.  Changes in a synaptic weight $\mathbf{W}_{ij}$ are governed by the learning rule $L(s)$, Eq. \eqref{stdprule}.  We begin by considering the total change in synaptic weight during an interval of length $T$ ms:
\begin{equation}
\Delta\mathbf{W}_{ij} = \mathbf{W}^0_{ij} \int_t^{t+T} \int_t^{t+T} L(t^{\prime\prime}-t^\prime)y_j(t^{\prime\prime})y_i(t^\prime) dt^{\prime\prime} dt^\prime
\end{equation}
where multiplying by the corresponding element of the adjacency matrix ensures that nonexistent synapses do not potentiate into existence.    Consider the trial-averaged rate of change:
\begin{equation}\label{dWdt1}
\frac{\langle \Delta\mathbf{W}_{ij}\rangle}{T} = \mathbf{W}^0_{ij}\frac{1}{T}\int_t^{t+T} \int_{t-t^\prime}^{t+T-t^\prime} L(s)\langle y_j(t^\prime+s)y_i(t^\prime) \rangle ds dt^\prime
\end{equation}
where $s=t^{\prime\prime}-t^\prime$ and $\langle \cdot \rangle$ denotes the trial average.  We first note that this contains the definition of the trial-averaged spike train cross-covariance:
 \begin{equation}
 \mathbf{C}_{ij}(s) = \frac{1}{T}\int_t^{t+T} \langle y_j(t^\prime+s)y_i(t^\prime) \rangle dt^\prime - r_ir_j
\end{equation}	
where $r_i$ is the time-averaged firing rate of neuron $i$ and subtracting off the product of the rates corrects for chance spike coincidences.  Inserting this definition into Eq. \eqref{dWdt1} yields:
\begin{equation} \label{dWdt2}
\frac{\langle \Delta\mathbf{W}_{ij}\rangle}{T} = \mathbf{W}^0_{ij} \int_{t-t^\prime}^{t+T-t^\prime} L(s)\left(r_ir_j + \mathbf{C}_{ij}(s) \right) ds 
\end{equation}

We then take the amplitude of individual changes in the synaptic weights to be small: $f_+, f_- < < W^\mathrm{max}$, where $\tau_\pm$ define the temporal shape of the STDP rule (see Eq. \eqref{stdprule}).  In this case, changes in the weights occur on a slower timescale than the width of the learning rule.  Taking $T >> \max{\left(\tau_+,\tau_-\right)}$ allows us to extend the limits of integration in Eq. \eqref{dWdt2} to $\pm\infty$, which gives Eq. \eqref{dWdt}.  Note that in the results we have dropped the angle brackets for convenience.  This can also be justified by the fact that the plasticity is self-averaging, since $\Delta \mathbf{W}_{ij}$ depends on the integrated changes over the period $T$.

%\begin{equation} \label{dWdt1}
%\frac{d\mathbf{W}_{ij}}{dt} = \mathbf{W}^0_{ij}\int_{-\infty}^\infty L(s) \langle y_j(t^\prime+s)y_i(t^\prime) \rangle ds
%\end{equation}
%Here, we have dropped the angle brackets around $\mathbf{W}_{ij}$ for convenience.  (Also note that when the synaptic plasticity is sufficiently slow, since $\Delta \mathbf{W}_{ij}$ depends on the integrated changes over the period $T$, the plasticity is self-averaging.)  

\subsection*{Spiking statistics}
\paragraph{}
In order to calculate $d\mathbf{W}_{ij}/dt$, we need to know the firing rates $r_i, r_j$ and spike train cross-covariance $\mathbf{C}_{ij}(s)$ (Eq. \eqref{dWdt}).  We take the weights to be constant on the fast timescale of $s$, so that the firing rates and spike train cross-covariances are stationary on that timescale.  We solve for the baseline firing rates in the network via the self-consistency relationship
\begin{equation*} \begin{aligned}
r_i &= r_i(\mu_i^{\mathrm{eff}}, \sigma) \mathrm{, where}\\
\mu_i^{\mathrm{eff}} &= \mu+\sum_j \left(\int_{-\infty}^\infty\mathbf{J}_{ij}(t)dt\right) \mathbf{W}_{ij}r_j
\end{aligned} \end{equation*}
for $i=1,\ldots,N$.  This gives the equilibrium state of each neuron's activity.  In order to calculate the spike train cross-covariances, we must consider temporal fluctuations around the baseline firing rates.

\paragraph{}
With sufficiently weak synapses compared to the background input, we can linearize each neuron's activity around the baseline state.  Rather than linearizing each neuron's firing rate around $r_i$, we follow \cite{doiron_oscillatory_2004, lindner_theory_2005, trousdale_impact_2012} and linearize each neuron's spike train around a realization of background activity, the uncoupled spike train $\mathbf{y}^0_i$ (Eq. \eqref{linear}).  The perturbation around the background activity is given by each neuron's linear response function, $\mathbf{A}_i(t)$, which measures the amplitude of firing rate fluctuations in response to perturbations of each neuron's input around the baseline $\mu_i^{\mathrm{eff}}$.  We calculate $\mathbf{A}(t)$ using standard methods based on Fokker-Planck theory for the distribution of a neuron's membrane potential \cite{richardson_firing-rate_2007, richardson_spike-train_2008}.

\paragraph{}
This yields Eq. \eqref{linear}, approximating a realization of each neuron's spike train as a mixed point and continuous process.  Spike trains are defined, however, as pure point processes.  Fortunately, Eq. \eqref{dWdt} shows that we do not need a prediction of individual spike train realizations, but rather of the trial-averaged spiking statistics.  We can solve Eq. \eqref{linear} for the spike trains in the frequency domain as:
\begin{equation*} \begin{aligned}
%\mathbf{y}(\omega) &= \mathbf{y}^0(\omega) + \left(\mathbf{W}\cdot \mathbf{K}(\omega)\right)\mathbf{y}(\omega) \\
\mathbf{y}(\omega) &= \left(\mathbf{I}- \left(\mathbf{W}\cdot \mathbf{K}(\omega)\right)\right)^{-1}\mathbf{y}^0(\omega)
\end{aligned} \end{equation*}
where as in the Results, $\mathbf{K}(\omega)$ is an interaction matrix defined by $\mathbf{K}_{ij}(\omega) = \mathbf{A}_i(\omega)\mathbf{J}_{ij}(\omega)$ and $\cdot$ denotes the element-wise product.  Averaging this expression over realizations of the background spike trains yields a linear equation for the instantaneous firing rates.  Averaging the spike trains $\mathbf{y}$ against each other yields the full cross-covariance matrix, Eq. \eqref{Cfull}.  It depends on the coupling strengths $\mathbf{W}$, the synaptic filters $\mathbf{J}_{ij}$ and neurons' linear response functions $\mathbf{A}$, and the covariance of the baseline spike trains, $\mathbf{C}^0$.

\paragraph{}
We can calculate the baseline covariance in the frequency domain, $\mathbf{C}^0(\omega) = \langle \mathbf{y}^0 \mathbf{y}^{0*}\rangle$, by first noting that it is a diagonal matrix containing each neuron's spike train power spectrum.  We calculate these using the renewal relationship between the spike train power spectrum $\mathbf{C}^0(\omega)$ and the first passage time density \cite{cox_point_1980}; the first passage time density for nonlinear integrate and fire models can be calculated using similar methods as for the linear response functions \cite{richardson_spike-train_2008}.  

\subsection*{Self-consistent theory for network plasticity}
\paragraph{}
We solve the system Eqs. \eqref{dWdt},\eqref{Cfull} for the evolution of each synaptic weight with the Euler method with a time step of $100$ seconds.  A package of code for solving the self-consistent theory and running the spiking simulations, in MATLAB and C, is available at http://sites.google.com/site/gabrielkochocker/code.  Additional code is available on request.

\subsection*{Derivation of motif dynamics}
\paragraph{}

The baseline structure of the network is defined by the adjacency matrix $\mathbf{W}^0$.  The frequencies of different motifs are:
\begin{equation}
\begin{aligned} \label{motifdef0}
p_0&=\frac{1}{N^2}\sum_{i,j}\mathbf{W}^0_{ij}, \\ 
q_0^\mathrm{div}&=\frac{1}{N^3}\sum_{i,j,k}\mathbf{W}^0_{ik}\mathbf{W}^0_{jk}-p_0^2,\\ 
q_0^\mathrm{con}&=\frac{1}{N^3}\sum_{i,j,k}\mathbf{W}^0_{ik}\mathbf{W}^0_{ij}-p_0^2, \\
q_0^\mathrm{ch}&=\frac{1}{N^3}\sum_{i,j,k}\mathbf{W}^0_{ij}\mathbf{W}^0_{jk}-p_0^2.\\
q_0^\mathrm{rec}&=\frac{1}{N^2}\sum_{i,j}\mathbf{W}^0_{ij}\mathbf{W}^0_{ji}-p_0^2.
\end{aligned}
\end{equation}
Each of the $q_0$ parameters refers to a different two-synapse motif.  In divergent motifs ($q_0^\mathrm{div}$), one neuron $k$ projects to two others, $i$ and $j$.  In convergent motifs ($q_0^\mathrm{con}$), two neurons $k$ and $j$ project to a third, $i$.  In chain motifs ($q_0^\mathrm{ch}$), neuron $k$ projects to neuron $j$, which projects to neuron $i$. Finally, in recurrent motifs ($q_0^\mathrm{rec}$) two neurons connect reciprocally.  In each of these equations, we subtract off $p_0^2$ to correct for the baseline frequencies expected in Erd\"{o}s-R\'{e}nyi random networks.  So, these parameters measure above-chance levels of motifs in the adjacency matrix $\mathbf{W}^0$.

We extend this motif definition to a weighted version, given by Eqs. \eqref{motifdef}.  Since our linear response theory for synaptic plasticity requires weak synapses, here we explicitly scale by the mean in-degree $\epsilon = \frac{1}{Np_0}$: 

\begin{equation}
\begin{aligned} \label{motifdef1}
\epsilon p&=\frac{1}{N^2}\sum_{i,j}\mathbf{W}_{ij}, \\ 
\epsilon^2 q^\mathrm{div}&=\frac{1}{N^3}\sum_{i,j,k}\mathbf{W}_{ik}\mathbf{W}_{jk}-\epsilon^2p^2,\\ 
\epsilon^2q^\mathrm{con}&=\frac{1}{N^3}\sum_{i,j,k}\mathbf{W}_{ik}\mathbf{W}_{ij}-\epsilon^2p^2, \\
\epsilon^2q^\mathrm{ch}&=\frac{1}{N^3}\sum_{i,j,k}\mathbf{W}_{ij}\mathbf{W}_{jk}-\epsilon^2p^2, \\
\epsilon q_\mathrm{X}^\mathrm{rec} &= \frac{1}{N^2}\sum_{i,j}\mathbf{W}_{ij}\mathbf{W}^0_{ji} - \epsilon pp_0, \\
\epsilon q_\mathrm{X}^\mathrm{div} &=  \frac{1}{N^3} \sum_{i,j,k}\mathbf{W}_{ik}\mathbf{W}^0_{jk}-\epsilon pp_0, \\
\epsilon q_\mathrm{X}^\mathrm{con} &= \frac{1}{N^3}\sum_{i,j,k}\mathbf{W}_{ik}\mathbf{W}^0_{ij}-\epsilon pp_0, \\
\epsilon q_\mathrm{X}^\mathrm{ch,A} &= \frac{1}{N^3}\sum_{i,j,k}\mathbf{W}_{ij}\mathbf{W}^0_{jk}-\epsilon pp_0, \\
\epsilon q_\mathrm{X}^\mathrm{ch,B} &= \frac{1}{N^3}\sum_{i,j,k}\mathbf{W}^0_{ij}\mathbf{W}_{jk}-\epsilon pp_0 \\
\end{aligned}
\end{equation}
Here we have defined the two-synapse motifs, as well as five auxiliary variables, $\left\{q_\mathrm{X}\right\}$.  These mixed motifs, defined by products of the weight and adjacency matrices, measure the strength of synapses \emph{conditioned} on their being part of a motif.  The motifs $\left\{q\right\}$, on the other hand, measure the total strength of the motifs.  While the variables $\left\{q_\mathrm{X}\right\}$ are not of direct interest, we will see that they are required in order to close the system of equations.  In comparison to the motif \emph{frequencies} $\left\{q_0 \right\}$, which measure motif frequencies in comparison to an independently \emph{connected} network, the motif \emph{strengths} are defined relative to an independently \emph{weighted} network.

\paragraph{}
We also scale the amplitude of individual synaptic changes, $L(s)$, by $\epsilon$.  We now go through the derivation of $dp/dt$, $dq^\mathrm{div}/dt$ and $dq_\mathrm{X}^\mathrm{div}/dt$ as examples; the other six variables follow the same steps.  First, note that the spike train cross-covariance matrix of the network, Eq. \eqref{Cfull}, can be expanded in the Fourier domain around the baseline covariance $\mathbf{C}^0(\omega)$:
\begin{equation} \label{Cseries}
\mathbf{C}(\omega) = \left(\sum_{i=0}^\infty \left(\mathbf{W}\cdot\mathbf{K}\right)^i\right)\mathbf{C}^0(\omega)\left(\sum_{j=0}^\infty\left(\left(\mathbf{W}\cdot\mathbf{K}\right)^*\right)^i\right)
\end{equation}
where the interaction matrix $\mathbf{W}\cdot\mathbf{K}$ is the element-wise product of the weight matrix $\mathbf{W}$ and the matrix of filters, $\mathbf{K}$.  Powers of $\mathbf{W}\cdot\mathbf{K}$ represent lengths of paths through the network.  Only taking into account up to length one paths yields (for $i\neq j$):
\begin{equation}
\mathbf{C}_{ij}(s) \approx \underbrace{\left(\mathbf{W}_{ij}\mathbf{K}_{ij}*\mathbf{C}^0_{jj}\right)(s)}_{\mathrm{forward connection}} + \underbrace{\left(\mathbf{C}^0_{ii}*\mathbf{W}_{ji}\mathbf{K}_{ji}^-\right)(s)}_{\mathrm{backward connection}} + \underbrace{\sum_{k}\left(\mathbf{W}_{ik}\mathbf{K}_{ik}*\mathbf{C}^0_{kk}*\mathbf{W}_{jk}\mathbf{K}_{jk}^- \right)(s)}_{\mathrm{common inputs}}.
\end{equation}
where we have inverse Fourier transformed for convenience in the following derivation and $\mathbf{K}^-(t)=\mathbf{K}(-t)$.

\paragraph{}
Differentiating each motif with respect to time, using the fast-slow STDP theory Eq. \eqref{dWdt} and inserting the first-order truncation of the cross-covariance functions, Eq. \eqref{Ctrunc}, yields:
\begin{equation} \begin{aligned} \label{dpdt1}
\epsilon \frac{dp}{dt} &= \frac{1}{N^2}\sum_{i,j} \mathbf{W}^0_{ij} \int_{-\infty}^\infty \epsilon L(s) \Big(r_ir_j + \delta_{ij}\mathbf{C}^0_{ij}(s) + \left(\mathbf{W}_{ij}\mathbf{K}_{ij}*\mathbf{C}^0_{jj}\right)(s) \\
&\;\;\; + \left(\mathbf{C}^0_{ii}*\mathbf{W}_{ji}\mathbf{K}_{ji}^-\right)(s) + \sum_{k}\left(\mathbf{W}_{ik}\mathbf{K}_{ik}*\mathbf{C}^0_{kk}*\mathbf{W}_{jk}\mathbf{K}_{jk}^- \right) \Big) ds
\end{aligned} \end{equation}
\begin{equation} \begin{aligned} \label{dqdivdt1}
\epsilon^2 \frac{dq^\mathrm{div}}{dt} &= \frac{2}{N^3} \sum_{i,j,k} \Big[\mathbf{W}_{ik}\mathbf{W}^0_{jk} \int_{-\infty}^\infty \epsilon L(s)\Big(r_jr_k + \delta_{jk}\mathbf{C}^0_{jk}(s) + \left(\mathbf{W}_{jk}\mathbf{K}_{jk}*\mathbf{C}^0_{kk}\right)(s) \\
&\;\;\; + \left(\mathbf{C}^0_{jj}*\mathbf{W}_{kj}\mathbf{K}_{kj}^-\right)(s) + \sum_{l}\left(\mathbf{W}_{jl}\mathbf{K}_{jl}*\mathbf{C}^0_{ll}*\mathbf{W}_{kl}\mathbf{K}_{kl}^- \right) \Big)ds \Big] - 2\epsilon^2 p\frac{dp}{dt}
\end{aligned} \end{equation}
\begin{equation} \begin{aligned} \label{dqxdivdt1}
\epsilon \frac{dq_\mathrm{X}^\mathrm{div}}{dt} &= \frac{1}{N^3}\sum_{i,j,k}\mathbf{W}^0_{jk} \mathbf{W}^0_{ik}\int_{-\infty}^\infty \epsilon L(s) \Big(r_ir_k +  \delta_{ik}\mathbf{C}^0_{ik}(s) + \left(\mathbf{W}_{ik}\mathbf{K}_{ik}*\mathbf{C}^0_{kk}\right)(s) \\
&\;\;\; + \left(\mathbf{C}^0_{ii}*\mathbf{W}_{ki}\mathbf{K}_{ki}^-\right)(s) + \sum_{l}\left(\mathbf{W}_{il}\mathbf{K}_{il}*\mathbf{C}^0_{ll}*\mathbf{W}_{kl}\mathbf{K}_{kl}^- \right) \Big)ds\Big] -\epsilon p_0\frac{dp}{dt}
\end{aligned} \end{equation}
We now assume that all neurons have the same firing rates, spike train autocovariances and linear response functions: $\forall i$, $r_i \equiv r$, $\mathbf{C}^0_{ii} \equiv C^0$ and $\mathbf{A}_i \equiv A$.  Since we model all postsynaptic currents with the same shape, this makes the matrix $\mathbf{K}$ a constant matrix; we replace its elements with the scalar $K$.  Also neglecting the weight bounds in $L(s)$ allows us to write:
\begin{equation}\begin{aligned} \label{dpdt2}
\frac{dp}{dt} &= r^2S \frac{1}{N^2}\sum_{i,j}\mathbf{W}^0_{ij} + S_F\frac{1}{N^2}\sum_{i,j}\mathbf{W}^0_{ij}\mathbf{W}_{ij} + S_B \frac{1}{N^2}\sum_{i,j}\mathbf{W}^0_{ij}\mathbf{W}_{ji} + S_C \frac{1}{N^2}\sum_{i,j,k} \mathbf{W}^0_{ij}\mathbf{W}_{ik}\mathbf{W}_{jk}
\end{aligned} \end{equation}
\begin{equation}\begin{aligned} \label{dqdivdt2}
\epsilon \frac{dq^\mathrm{div}}{dt} &= r^2 S \frac{2}{N^3}\sum_{i,j,k}\mathbf{W}_{ik}\mathbf{W}^0_{jk} + S_F \frac{2}{N^3}\sum_{i,j,k}\mathbf{W}_{ik}\mathbf{W}^0_{jk}\mathbf{W}_{jk} \\
\;\;\; &+ S_B\frac{2}{N^3}\sum_{i,j,k}\mathbf{W}_{ik}\mathbf{W}^0_{jk}\mathbf{W}_{kj} + S_C\frac{2}{N^3}\sum_{i,j,k,l}\mathbf{W}_{ik}\mathbf{W}^0_{jk}\mathbf{W}_{jl}\mathbf{W}_{kl} - 2\epsilon p\frac{dp}{dt}
\end{aligned} \end{equation}
\begin{equation}\begin{aligned} \label{dqxdivdt2}
\frac{dq_\mathrm{X}^\mathrm{div}}{dt} &= r^2S\frac{1}{N^3}\sum_{i,j,k} \mathbf{W}^0_{jk}\mathbf{W}^0_{ik} + S_F\frac{1}{N^3}\sum_{i,j,k}\mathbf{W}^0_{jk}\mathbf{W}^0_{ik}\mathbf{W}_{ik} \\
\;\;\; &+ S_B\frac{1}{N^3}\sum_{i,j,k}\mathbf{W}^0_{jk}\mathbf{W}^0_{ik}\mathbf{W}_{ki} + S_C\frac{1}{N^3}\sum_{i,j,k,l} \mathbf{W}^0_{jk}\mathbf{W}^0_{ik}\mathbf{W}_{il}\mathbf{W}_{kl} -  p_0\frac{dp}{dt}
\end{aligned} \end{equation}
where we have cancelled off an $\epsilon$ from the left and right-hand sides.  We have absorbed the integrals over the STDP rule and the spiking covariances into $r^2S$, $S_F$, $S_B$ and $S_C$.  These correspond, respectively, to the total STDP-weighted spiking covariances from chance coincidence, forward connections, backward connections, and common input:
\begin{equation}\begin{aligned} 
S &= \int_{-\infty}^\infty L(s) \, ds \label{Sdef} \\ 
\end{aligned} \end{equation}
\begin{equation}\begin{aligned} 
S_F &= \int_{-\infty}^\infty L(s)\left(K(t)*C^0(s) \right)ds \label{SFdef} \\
\end{aligned} \end{equation}
\begin{equation}\begin{aligned} 
S_B &= \int_{-\infty}^\infty L(s)\left(C^0(s)*K^-(t) \right)ds \label{SBdef} \\
\end{aligned} \end{equation}
\begin{equation}\begin{aligned} 
S_C &= \int_{-\infty}^\infty L(s)\left(K(t)*C^0(s)*K^-(t) \right)ds \label{SCdef}
\end{aligned}\end{equation}
These parameters depend on the spike train auto-covariance $C^0(s)$ and linear response functions $A(t)$ of neurons.  These functions can change as the network'Õs operating point does; for instance, as the leak reversal $V_L$ increases, neurons will shift to a mean-driven, more oscillatory spiking regime and $C^0(s)$ and $K(t;A)$ will change.  As the mean synaptic weight changes, the firing rates will change and this can also affect $C^0(s)$ and $K(t;A)$.  We have assumed weak synapses, so we will fix these at their value at $p = p_0W^\mathrm{max}/2$.  Without this approximation, the system is transcendental.

\paragraph{}
Each dynamical equation now contains four different sums of products of the weight and adjacency matrices.  First examining $dp/dt$, we see that the first three sums correspond to defined motifs: $1/N^2 \sum_{i,j}\mathbf{W}^0_{ij} = p_0$, $1/N^2 \sum_{i,j}\mathbf{W}^0_{ij}\mathbf{W}_{ij} = p$ and $1/N^2 \sum_{i,j}\mathbf{W}^0_{ij}\mathbf{W}_{ji} = q_\mathrm{X}^\mathrm{rec} + pp_0$.  The last term in Eq. \eqref{dpdt2}, however, corresponds to a third-order motif mixed between the weight and adjacency matrices.  Similarly, third- and fourth-order mixed motifs appear in Eqs. \ref{dqdivdt2} and \ref{dqxdivdt2}.  In order to calculate these, we extend a re-summing technique developed in \cite{hu_motif_2013}.  We assume that there are no third- or higher-order correlations between elements of the weight and/or adjacency matrices, and approximate the frequency of each of these higher-order motifs by the number of ways it can be composed of one and two-synapse motifs.  For a third order motif, this corresponds to adding up the likelihoods that all three synapses occur by chance and that each possible combination of one synapse and a two-synapse motif occur.  In Eq. \eqref{dpdt2},
\begin{equation} \label{resum1}
\sum_{i,j,k}\mathbf{W}^0_{ij}\mathbf{W}_{ik}\mathbf{W}_{jk} \approx \epsilon^2N^3 \left(p_0\left(q^\mathrm{div}+p^2\right) + p\left(q_\mathrm{X}^\mathrm{con} + q_\mathrm{X}^\mathrm{ch,B}\right)\right).
\end{equation}
and for the four-synapse motif in Eq. \eqref{dqdivdt2},
\begin{equation} \label{resum2}
\sum_{i,j,k,l}\mathbf{W}_{ik}\mathbf{W}^0_{jk}\mathbf{W}_{jl}\mathbf{W}_{kl} \approx \epsilon^3N^4 \left(p^3p_0+p^2\left(q_\mathrm{X}^\mathrm{div}+q_\mathrm{X}^\mathrm{con}+q_\mathrm{X}^\mathrm{ch,B}\right)+pp_0\left(q^\mathrm{div}+q^\mathrm{ch}\right)+q^\mathrm{div}q_\mathrm{X}^\mathrm{div}+q^\mathrm{ch}q_\mathrm{X}^\mathrm{con} \right)
\end{equation}

\paragraph{}
This re-summing, along with the inclusion of the mixed motifs $\left\{q_\mathrm{X}\right\}$, is what allows us to close the motif dynamics.  Re-summing each third- and fourth-order motif in our system in terms of two-synapse motifs yields, after simplification, the final motif dynamics:
\begin{equation} \begin{aligned} \label{dpdtfull}
\frac{dp}{dt} &= p_0r^2S + \epsilon \left[pS_F + \left(q_\mathrm{X}^\mathrm{rec}+p_0p\right)S_B + \frac{1}{p_0}\left(p_0\left(q^\mathrm{div}+p^2\right)+p\left(q_\mathrm{X}^\mathrm{con}+q_\mathrm{X}^\mathrm{ch,B}\right) \right)S_C\right]
\end{aligned} \end{equation}
\begin{equation}\begin{aligned} \label{dqdivdtfull}
\frac{dq^\mathrm{div}}{dt} &= 2r^2Sq_\mathrm{X}^\mathrm{div} + 2\epsilon\left[q^\mathrm{div}S_F + \left(p_0q^\mathrm{ch}+pq_\mathrm{X}^\mathrm{div}\right)S_B + \frac{1}{p_0}\left(q^\mathrm{ch}\left(q_\mathrm{X}^\mathrm{con}+pp_0\right) + q_\mathrm{X}^\mathrm{div}\left(q^\mathrm{div}+p^2\right) \right)S_C \right]
\end{aligned} \end{equation}
\begin{equation}\begin{aligned} \label{dqcondtfull}
\frac{dq^\mathrm{con}}{dt} &= 2r^2S q_\mathrm{X}^\mathrm{con} + 2\epsilon\left[q^\mathrm{con}S_F + \left(p_0q^\mathrm{ch}+pq_\mathrm{X}^\mathrm{con}\right)S_B + \frac{1}{p_0}\left(q^\mathrm{con}\left(q_\mathrm{X}^\mathrm{ch,B}+pp_0\right) + q_\mathrm{X}^\mathrm{con}\left(q^\mathrm{div}+p^2\right) \right)S_C \right]
\end{aligned} \end{equation}
\begin{equation}\begin{aligned} \label{dqchdtfull}
\frac{dq^\mathrm{ch}}{dt} &= r^2S\left(q_\mathrm{X}^\mathrm{ch,A}+q_\mathrm{X}^\mathrm{ch,B}\right) + \epsilon \Big[2q^\mathrm{ch}S_F + \left( p_0\left(q^\mathrm{con} + q^\mathrm{div}\right) + p\left(q_\mathrm{X}^\mathrm{ch,A}+q_\mathrm{X}^\mathrm{ch,B}\right)\right)S_B \\
&\;\;\; + \frac{1}{p_0}\left(q_\mathrm{X}^\mathrm{ch,A}\left(q^\mathrm{div}+p^2\right)+q_\mathrm{X}^\mathrm{ch,B}\left(q^\mathrm{div}+q^\mathrm{con}+p^2\right) \right)S_C \Big]
\end{aligned} \end{equation}
\begin{equation}\begin{aligned} \label{dqxrecdtfull}
\frac{dq_\mathrm{X}^\mathrm{rec}}{dt} &= r^2Sq_0^\mathrm{rec} + \epsilon\Big[q_\mathrm{X}^\mathrm{rec}S_F + \left(1-p_0\right)\left(q_\mathrm{X}^\mathrm{rec}+pp_0 \right)S_B \\
&\;\;\; + \frac{1}{p_0}\left(q_0^\mathrm{rec}\left(q^\mathrm{div}+p^2\right) + q_\mathrm{X}^\mathrm{ch,B}\left(q_\mathrm{X}^\mathrm{ch,B}+pp_0\right) + q_\mathrm{X}^\mathrm{con}\left(q_\mathrm{X}^\mathrm{con}+pp_0\right) \right)S_C \Big]
\end{aligned} \end{equation}
\begin{equation}\begin{aligned} \label{dqxdivdtfull}
\frac{dq_\mathrm{X}^\mathrm{div}}{dt} &= r^2Sq_0^\mathrm{div} + \epsilon\left[q_\mathrm{X}^\mathrm{div}S_F + \left(pq_0^\mathrm{div}+p_0q_\mathrm{X}^\mathrm{ch,B}\right)S_B +\frac{1}{p_0}\left(q_0^\mathrm{div}\left(q^\mathrm{div}+p^2 \right) + q_\mathrm{X}^\mathrm{ch,B}\left(q_\mathrm{X}^\mathrm{con}+pp_0 \right) \right)S_C \right]
\end{aligned} \end{equation}
\begin{equation}\begin{aligned} \label{dqxcondtfull}
\frac{dq_\mathrm{X}^\mathrm{con}}{dt} &= r^2Sq_0^\mathrm{con} + \epsilon\left[q_\mathrm{X}^\mathrm{con}S_F + \left(pq_0^\mathrm{con}+p_0q_\mathrm{X}^\mathrm{ch,A} \right)S_B + \frac{1}{p_0}\left(q_0^\mathrm{con}\left(q^\mathrm{div}+p^2\right) + q_\mathrm{X}^\mathrm{con}\left(q_\mathrm{X}^\mathrm{ch,B}+pp_0\right) \right)S_C\right]
\end{aligned} \end{equation}
\begin{equation}\begin{aligned} \label{dqxchAdtfull}
\frac{dq_\mathrm{X}^\mathrm{ch,A}}{dt} &= r^2Sq_0^\mathrm{ch} + \epsilon \left[q_\mathrm{X}^\mathrm{ch,A}S_F + \left(pq_0^\mathrm{ch}+p_0q_\mathrm{X}^\mathrm{con} \right)S_B+\frac{1}{p_0}\left(q_0^\mathrm{ch}\left(q^\mathrm{div}+p^2\right) + q_\mathrm{X}^\mathrm{con}\left(q_\mathrm{X}^\mathrm{con}+pp_0\right) \right)S_C \right]
\end{aligned} \end{equation}
\begin{equation}\begin{aligned} \label{dqxchBdtfull}
\frac{dq_\mathrm{X}^\mathrm{ch,B}}{dt} &= r^2 S q_0^\mathrm{ch} + \epsilon \Big[q_\mathrm{X}^\mathrm{ch,B}S_F + \left(pq_0^\mathrm{ch}+p_0q_\mathrm{X}^\mathrm{div}\right)S_B + \frac{1}{p_0}\left(q_0^\mathrm{ch}\left(q^\mathrm{div}+p^2\right) + q_\mathrm{X}^\mathrm{ch,B}\left(q_\mathrm{X}^\mathrm{ch,B}+pp_0\right) \right)S_C \Big]
\end{aligned} \end{equation}

\paragraph{}
Examination of these equations reveals how different types of joint spiking activity affect motif dynamics.  Chance spiking coincidence (the $r^2S$ terms) couple each motif to the mixed version of itself, and each mixed motif to the baseline structure of the adjacency matrix.  With Hebbian STDP and excitatory synapses, $S_F > 0$ and $S_B<0$.  So, spiking covariance from forward connections provide positive feedback, reinforcing the current network structure.  Spiking covariance from backward connections and common input couple divergent, convergent and chain motifs to each other.

\paragraph{}
The dynamics on the invariant set (Results: Balanced STDP of the mean synaptic weight, Fig. 6) were plotted in MATLAB.  The vector fields of Figs. 8 and 9 were calculated in XPPAUT.  For those figures, results from simulations of the full spiking network were plotted in MATLAB and then overlaid on the vector fields from XPPAUT.

\subsubsection*{Plasticity of loops and feedforward chains}
The chain variable $q^\mathrm{ch}$ includes both feedforward and recurrent loops.  (Feedforward chains correspond to $k\neq i$ in the definition of $q^\mathrm{ch}$, Eq. \eqref{motifdef1}, and recurrent loops to $k = i$.)  As in the main text, we break $q^\mathrm{ch}$ into these two cases: $q^\mathrm{ch} = q^\mathrm{rec} + q^\mathrm{ff}$, where
\begin{equation} \begin{aligned}
\epsilon^2 q^\mathrm{rec} &= \frac{1}{N^3}\sum_{i,j,k}\delta_{ik}\mathbf{W}_{ij}\mathbf{W}_{jk} = \frac{1}{N^3}\sum_{i,j} \mathbf{W}_{ij}\mathbf{W}_{ji} \\
\epsilon^2 q^\mathrm{ff} &= \frac{1}{N^3}\sum_{i,j,k}\left(1-\delta_{ik} \right)\mathbf{W}_{ij}\mathbf{W}_{jk}-\epsilon^2p^2
\end{aligned} \end{equation}
We also define an auxiliary variable which we will require in the dynamics of $q^\mathrm{rec}$:
\begin{equation} \label{qx2rec}
\epsilon^2 q_\mathrm{X2}^\mathrm{rec} = \frac{1}{N^3}\sum_{i,j}\mathbf{W}_{ij}^2\mathbf{W}^0_{ji}
\end{equation}
which is proportional to the conditioned second moment of weights that are part of disynaptic loops.  The dynamics of $q^\mathrm{rec}$ are calculated exactly as for the other motifs and are:
\begin{equation}
\frac{1}{2\epsilon}\frac{dq^\mathrm{rec}}{dt} = r^2 Sp_0\left(q_\mathrm{X}^\mathrm{rec} + pp_0\right)+S_Fq^\mathrm{rec}+S_Bq_\mathrm{X2}^\mathrm{rec}
\end{equation}
where the new auxiliary variable obeys
\begin{equation} \label{dqX2recdt}
\frac{1}{2\epsilon}\frac{dq_\mathrm{X2}^\mathrm{rec}}{dt} = r^2Sp_0\left(q_\mathrm{X}^\mathrm{rec} +pp_0\right)+S_Fq_\mathrm{X2}^\mathrm{rec}+S_Bq^\mathrm{rec}
\end{equation}
We can then recover the dynamics of feedforward chains as:
\begin{equation} \begin{aligned} \label{dqffdt}
\frac{dq^\mathrm{ff}}{dt} &= \frac{dq^\mathrm{ch}}{dt} - \frac{dq^\mathrm{rec}}{dt} \\
&= r^2S\left(q_\mathrm{X}^\mathrm{ch,A}+q_\mathrm{X}^\mathrm{ch,B}\right) + \epsilon \Big[ -2r^2 Sp_0\left(q_\mathrm{X}^\mathrm{rec} + pp_0\right) + 2S_F\left(q^\mathrm{ch}-q^\mathrm{rec}\right) \\
&\;\;\; + \left(p_0\left(q^\mathrm{con} + q^\mathrm{div}\right) + p\left(q_\mathrm{X}^\mathrm{ch,A}+q_\mathrm{X}^\mathrm{ch,B}\right)-2q_\mathrm{X2}^\mathrm{rec}\right)S_B \\
&\;\;\; + \frac{1}{p_0}\left(q_\mathrm{X}^\mathrm{ch,A}\left(q^\mathrm{div}+p^2\right)+q_\mathrm{X}^\mathrm{ch,B}\left(q^\mathrm{div}+q^\mathrm{con}+p^2\right) \right)S_C \Big]
\end{aligned} \end{equation}

\subsubsection*{Unbalanced STDP}
\paragraph{}
When there is an imbalance between the net amounts of potentiation and depression in the STDP rule, the motif dynamics are governed by simpler equations.  If $S \sim \mathcal{O}(1)$, the $\mathcal{O}(\epsilon)$ terms in Eqs. \ref{dpdtfull}-\ref{dqxchBdtfull} are negligible.  For each mixed motif,
\begin{equation}
q_\mathrm{X}(t) = r^2Sq_0 t + q_\mathrm{X}(0)
\end{equation}
so that 
\begin{align} \label{unbal}
p(t) &= p_0r^2S t + p(0) \\
q^\mathrm{div}(t) &= q^\mathrm{div}(0) + q_\mathrm{X}^\mathrm{div}(0)r^2St + \frac{1}{2}q_0^\mathrm{div} \left(r^2S\right)^2 t^2 \\
q^\mathrm{con}(t) &= q^\mathrm{con}(0) + q_\mathrm{X}^\mathrm{con}(0)r^2St +  \frac{1}{2}q_0^\mathrm{con} \left(r^2S\right)^2 t^2 \\
q^\mathrm{ch}(t) &= q^\mathrm{ch}(0) + \left(q_\mathrm{X}^\mathrm{ch,A}(0) + q_\mathrm{X}^\mathrm{ch,B}(0)\right)r^2St + \frac{1}{2}q_0^\mathrm{ch} \left(r^2S\right)^2 t^2
\end{align}
Writing $q_\mathrm{X}^\mathrm{ch} = q_\mathrm{X}^\mathrm{ch,A} + q_\mathrm{X}^\mathrm{ch,B}$ puts the dynamics for all the motifs in the same form.  The motifs expand from the initial conditions and baseline structure of the network.  Note that since the quadratic term is proportional to $S^2$, even when STDP is depression-dominated the long-term dynamics are expansive rather than contractive.

\section*{Acknowledgments}
We thank Kre\v{s}imir Josi\'{c} and the members of the Doiron group for useful comments on the manuscript.  Funding is provided by NSF-DMS1313225 (B.D.).

%\section*{References}
\bibliographystyle{plos2009}
\bibliography{STDP_LR}
% Either type in your references using
% \begin{thebibliography}{}
% \bibitem{}
% Text
% \end{thebibliography}
%
% OR
%
% Compile your BiBTeX database using our plos2009.bst
% style file and paste the contents of your .bbl file
% here.
% 

%\newpage
%\section*{Figures}
% This section is for figure legends only, do not include
% graphics in your manuscript file.
%

\newpage
\section*{Tables}
% 
% See introductory notes if you wish to include sideways tables.
%
% NOTE: Please look over our table guidelines at http://www.plosone.org/static/figureGuidelines#tables to make sure that your tables meet our requirements. Certain types of spacing, cell merging, and other formatting tricks may have unintended results and will be returned for revision.
%

\begin{table}[!ht]
\caption{
\bf{Model parameters}}
\begin{tabular}{|l|c|r|}
\hline
Parameter & Description & Value \\ \hline
$C$ & Membrane capacitance & 1 $\mu\mathrm{F/cm}^2$ \\ \hline
$g_L$ & Leak conductance & $0.1\mathrm{ mS/cm}^2$ \\ \hline
$V_L$ & Leak reversal potential & -72 mV \\ \hline
$\Delta$ & Action potential steepness & 1.4 mV\\ \hline
$V_T$ & Action potential initiation threshold & -48 mV \\ \hline
$V_{th}$ & Action potential threshold & 30 mV \\ \hline
$V_{re}$ & Action potential reset & -72 mV \\ \hline
$\tau_{ref}$ & Action potential width & 2 ms \\ \hline
$\mu$ & External input mean & 1 $\mu\mathrm{A/cm}^2$ \\ \hline
$\sigma$ & External input standard deviation & 9 mV \\ \hline
$N$ & Number of neurons & 1000 \\ \hline
$p_0$ & Connection density & .15 \\ \hline
$W^\mathrm{max}$ & Maximum synaptic weight & $5 \, \mu\mathrm{A/cm}^2$ \\ \hline
$\tau_\mathrm{S}$ & Synaptic time constant & 5 ms \\ \hline
\end{tabular}
\begin{flushleft}
\end{flushleft}
\label{tab:label}
 \end{table}

\newpage
%\section*{Figures}
\begin{figure}[ht!]
	{\center{\includegraphics{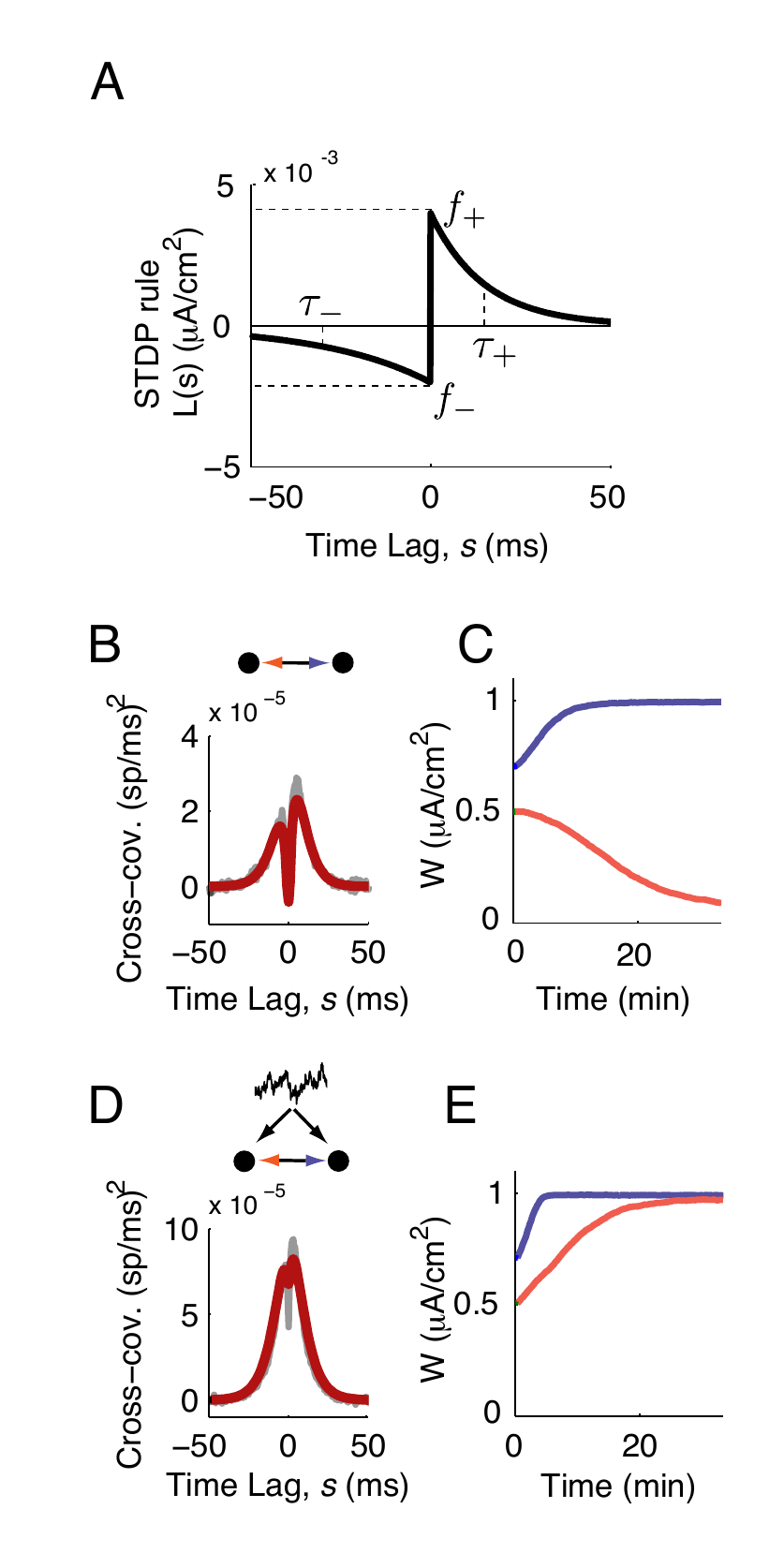} } }
\caption{
	{\bf Network structure shapes synaptic plasticity.}  (A) The STDP rule, $L(s)$, is composed of exponential windows for depression (-) and potentiation (+).  Each is defined by its amplitude $f_\pm$ and timescale $\tau_\pm$.  (B) Spike train cross-covariance function for a pair of neurons with no common input, so that synapses between the two neurons are the only source of spiking covariance.  Shaded lines: simulation, solid lines: theory (Eq. \eqref{Cfull}).  (C,E) Synaptic weight (peak EPSC amplitude) as a function of time in the absence (C) and presence (E) of common input.  (D) Spike train cross-covariance function for a pair of neurons with common input, $c=0.05$.  Common input was modeled as an Ornstein-Uhlenbeck process with a 5 ms timescale.
}
\label{Fig1}
\end{figure}

%\newpage

\begin{figure}[ht!]
{\center{\includegraphics{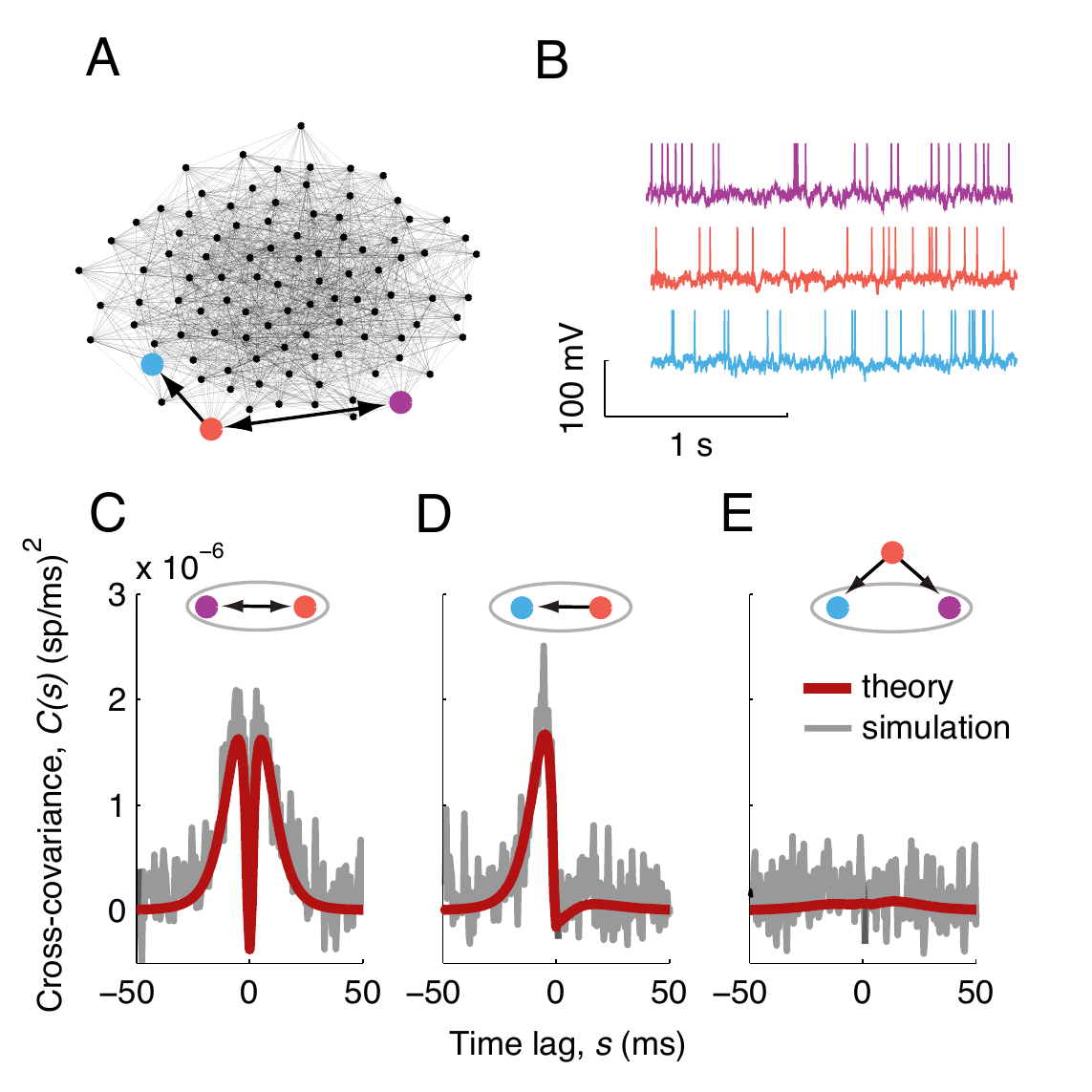} } }
\caption{
	{\bf Linear response theory for spiking covariances.} (A)  Illustration of the network connectivity for a subset of 100 neurons.  Three neurons, and the connections between them, are highlighted.  Nodes are positioned by the Fruchterman-Reingold force algorithm. (B) Example voltage traces for the three highlighted neurons.  (C-E) Spike train cross-covariance functions for the three combinations of labeled neurons.  Top: A shaded ellipse contains the pair of neurons whose cross-covariance is shown.  Shaded lines: simulations, red lines: linear response theory.  }
\label{Fig2}
\end{figure}

%\newpage

\begin{figure}[ht!]
	{\center{\includegraphics{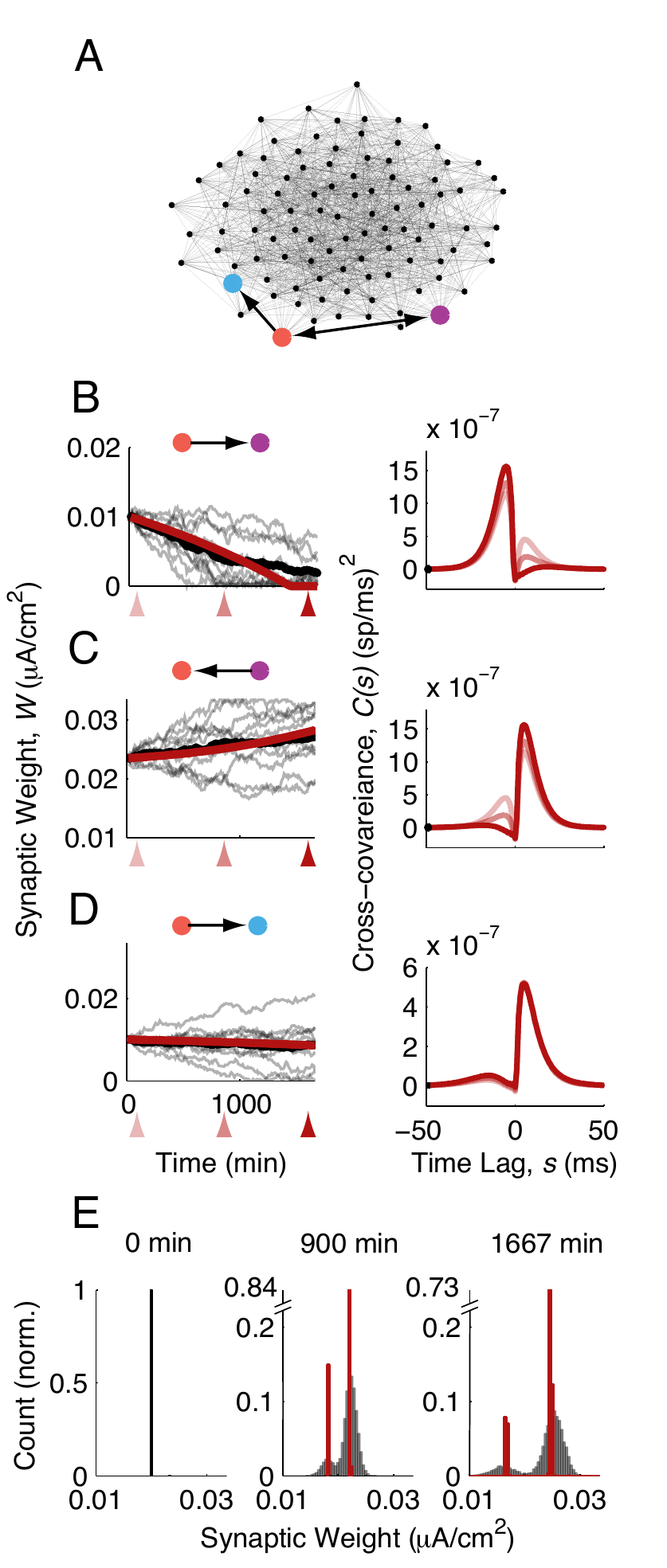} } }
\caption{
{\bf STDP in recurrent networks with internally generated spiking covariance.}  (A) As in Fig. \ref{Fig2}A.  (B-D) Left, Synaptic weight versus time for each of the three synapses in the highlighted network.  Shaded lines: simulation, individual trials.  Solid black lines: simulation, trial-average.  Solid red lines: theory.  Right, spike train cross-covariances at the three time points marked on the left (linear response theory).  (E) Histogram of synaptic weights at three time points.  Red, theory.  Shaded: simulation.
}
\label{Fig3}
\end{figure}

%\newpage

\begin{figure}[ht!]
{\center{\includegraphics{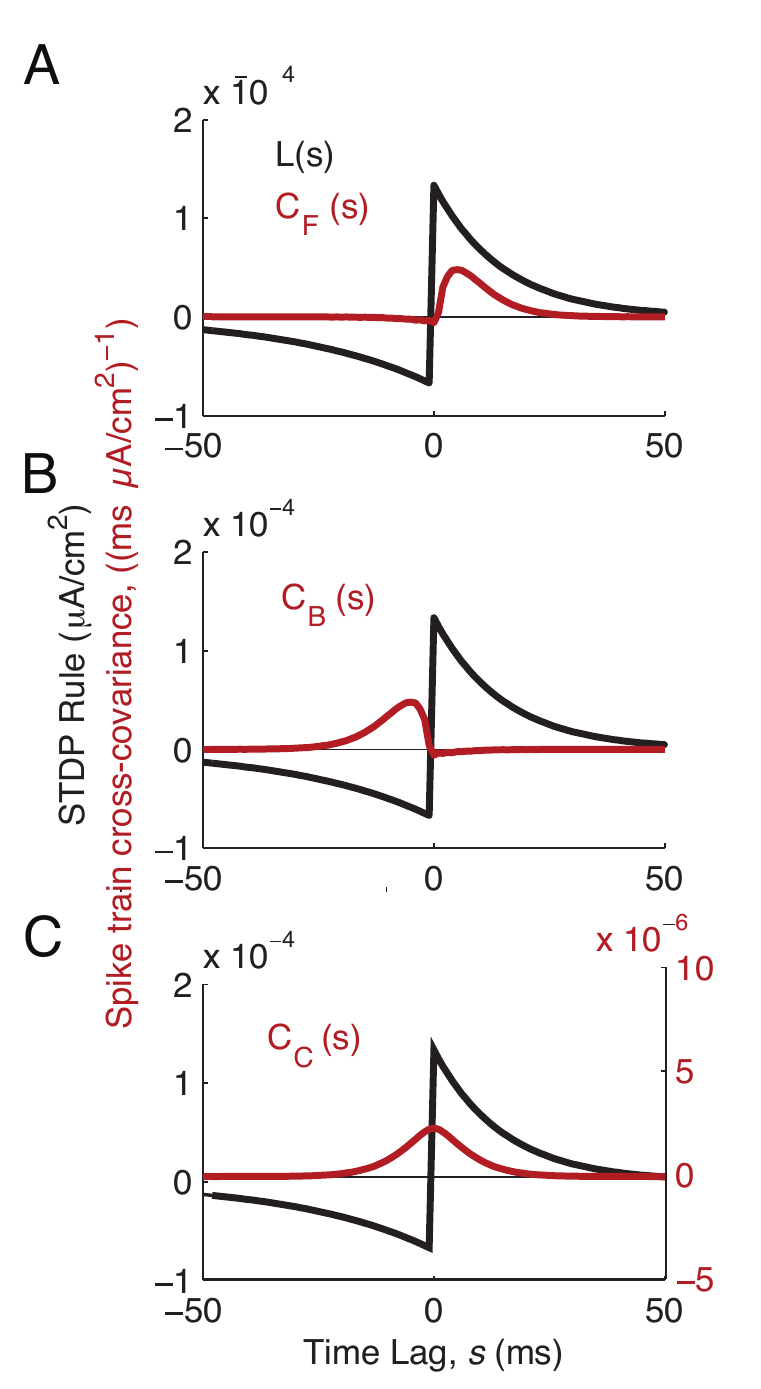} } }
\caption{
{\bf Different sources of spiking covariance interact with different parts of the STDP rule.}  Black: STDP rule.  Red: spike train cross-covariances, from Eq. \eqref{Ctrunc}. (A) Covariance from forward connections interacts with the potentiation side of the STDP rule.  (B) Covariance from backward connections interacts with the depression side of the STDP rule.  (C) Covariance from common input is temporally symmetric and interacts with both the potentiation and depression sides of the STDP rule.}
\label{Fig4}
\end{figure}

%\newpage

\begin{figure}[ht!]
{\center{\includegraphics{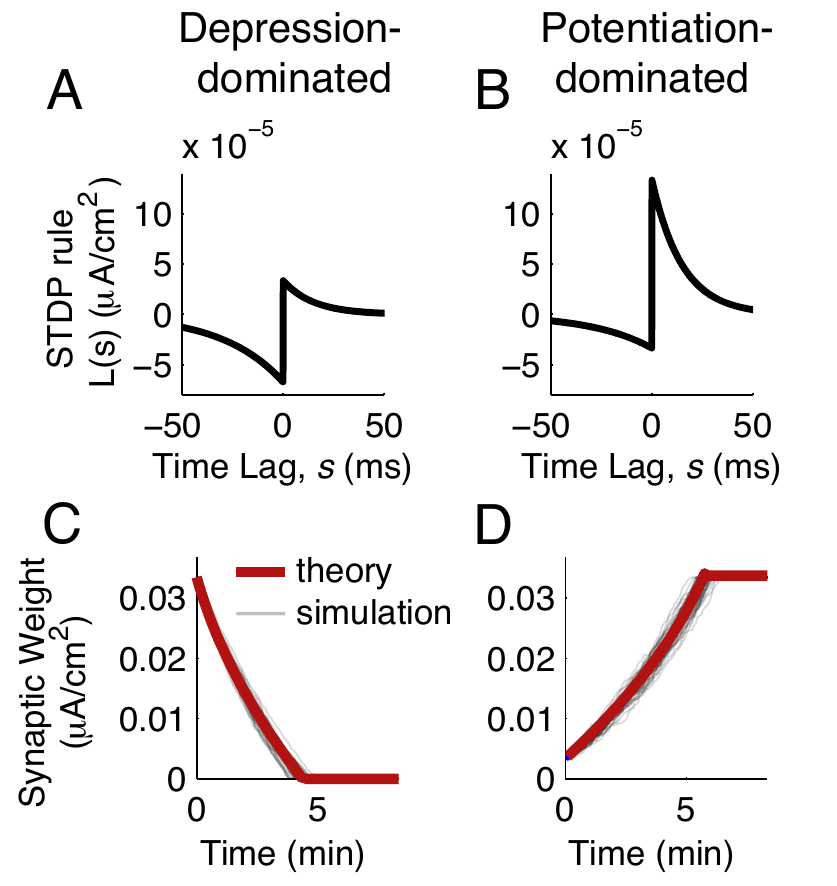} } }
\caption{
{\bf Unbalanced plasticity gives rise to simple weight dynamics.}  (A) A depression-dominated STDP rule: the amount of depression (integral of the depression side of the curve) is twice the amount of potentiation.  (B) A potentiation-dominated STDP rule: the amount of potentiation is twice the amount of depression.  (C) Evolution of synaptic weights with depression-dominated STDP: all weights depress.  (D) Evolution of synaptic weights with potentiation-dominated STDP: all weights potentiate.  Red lines: theory for mean synaptic weight.  Shaded lines: simulation of individual synaptic weights.
}
\label{Fig5}
\end{figure}

%\newpage

\begin{figure}[ht!]
{\center{\includegraphics{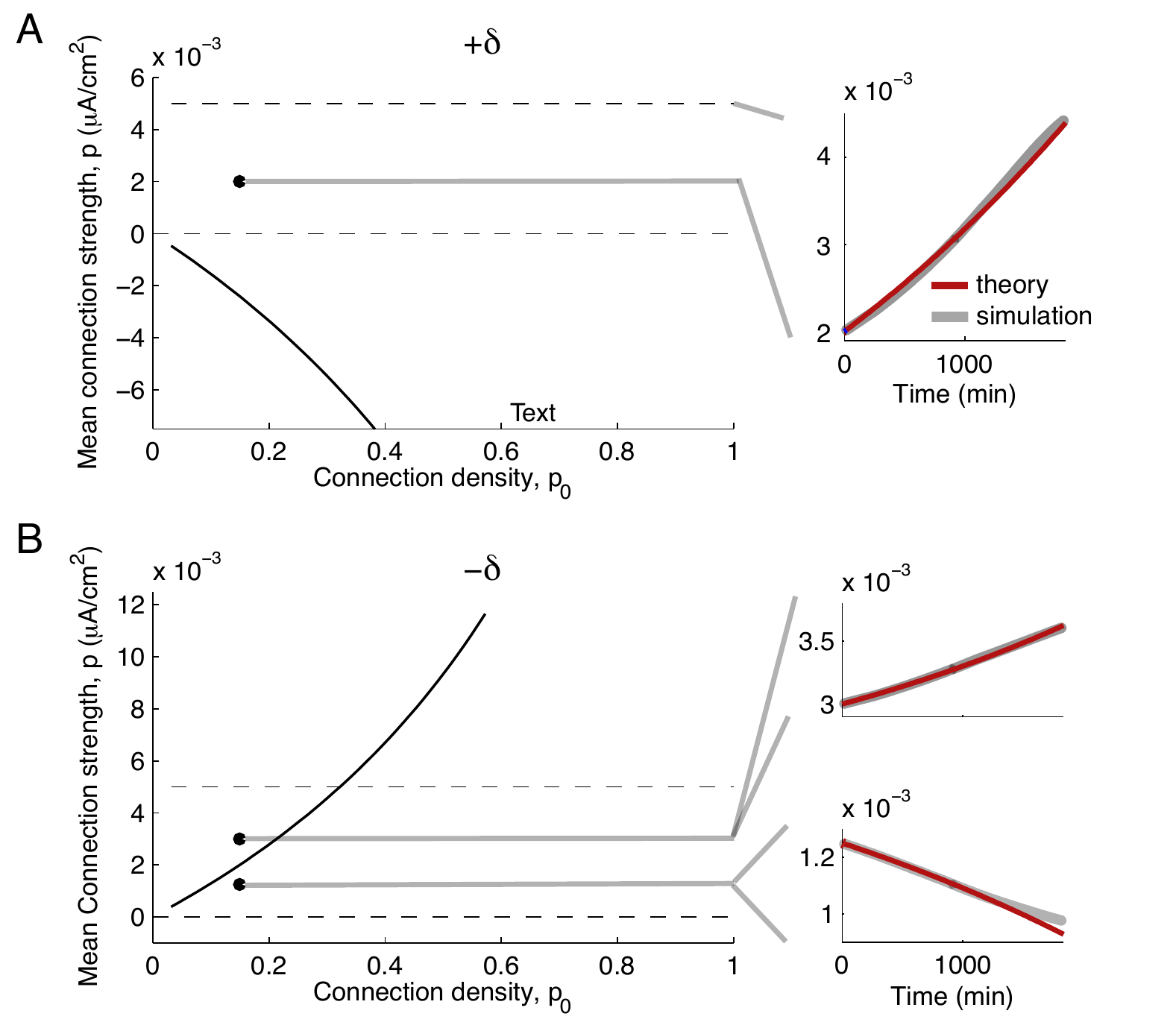} } }
\caption{
	{\bf Balanced plasticity of the mean synaptic weight.}    
%	When STDP is balanced between potentiation and depression, the dynamics on the Erd\"{o}s-R\'{e}nyi invariant set are quadratic in $p$ but only one of the two fixed points (the unstable one) is near the bounds $\left[0,p_0W^\mathrm{max}\right]$ \alkcomment{Is this sentence referring to something that's not in the figure?  If so it should be removed} 
	(A) When the STDP rule is balanced and potentiation-dominated, the unstable fixed point for $p$ is negative and decreases with the connection probability. So, the mean synaptic weight always increases. (B) When the STDP rule is balanced and depression-dominated, the unstable fixed point is positive and increases with the connection probability.  (A,B) Left: Dashed lines mark bounds for the mean synaptic weight, at $0$ and $p_0W^\mathrm{max}$.  Black curves track the location of the unstable fixed point of $p$ as the connection probability, $p_0$, varies.  Black dots mark initial conditions for the right panels.  (A,B) Right: Dynamics of the mean synaptic weight in each of the regimes of the left plots.  Red lines mark the reduced theory's prediction (Eq. \eqref{dpdt_1d}) and shaded lines the result of simulating the full spiking network. 
}
\label{Fig6}
\end{figure}

%\newpage

\begin{figure}[ht!]
{\center{\includegraphics{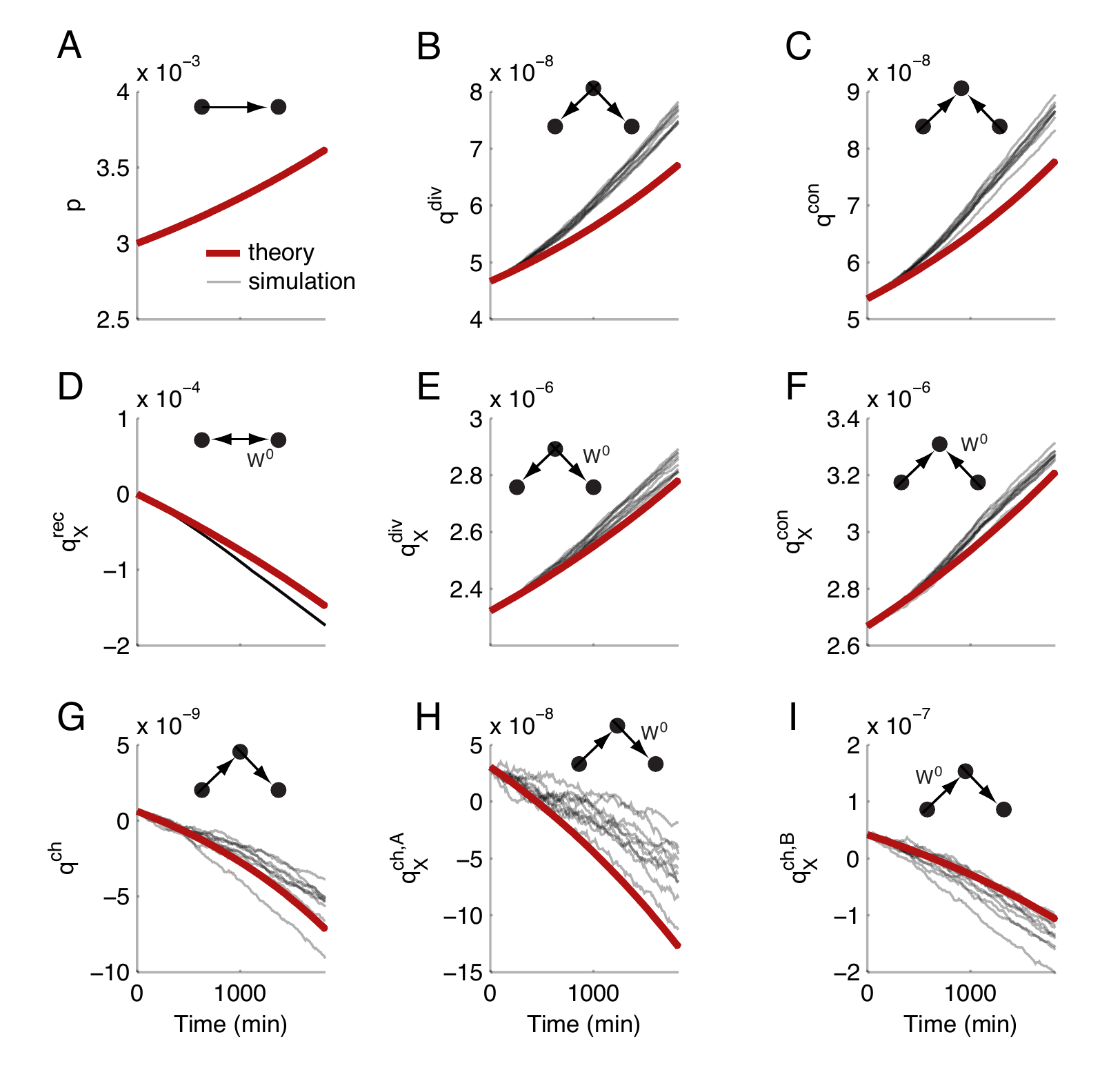} } }
\caption{
{\bf Reduced theory for the plasticity of two-synapse motifs.}  In each panel, the strength of a different motif or mixed motif is plotted as it evolves.  (A) Mean synaptic weight.  (B) Divergent motifs.  (C) Convergent motifs.  (D) Mixed recurrent motifs (strength of connections conditioned on their being part of a two-synapse loop).  (E) Mixed divergent motifs (strength of individual synapses conditioned on their being part of a divergent motif).  (F) Mixed convergent motifs.  (G) Chain motifs.  (H) Mixed chains type A (strength of individual synapses conditioned on their being the first in a chain).  (I) Mixed chains type B (strength of individual synapses conditioned on their being the second in a chain).  The STDP rule was in the depression-dominated balanced regime, as in Fig. 7B.
}
\label{Fig7}
\end{figure}

%\newpage

\begin{figure}[ht!]
{\center{\includegraphics{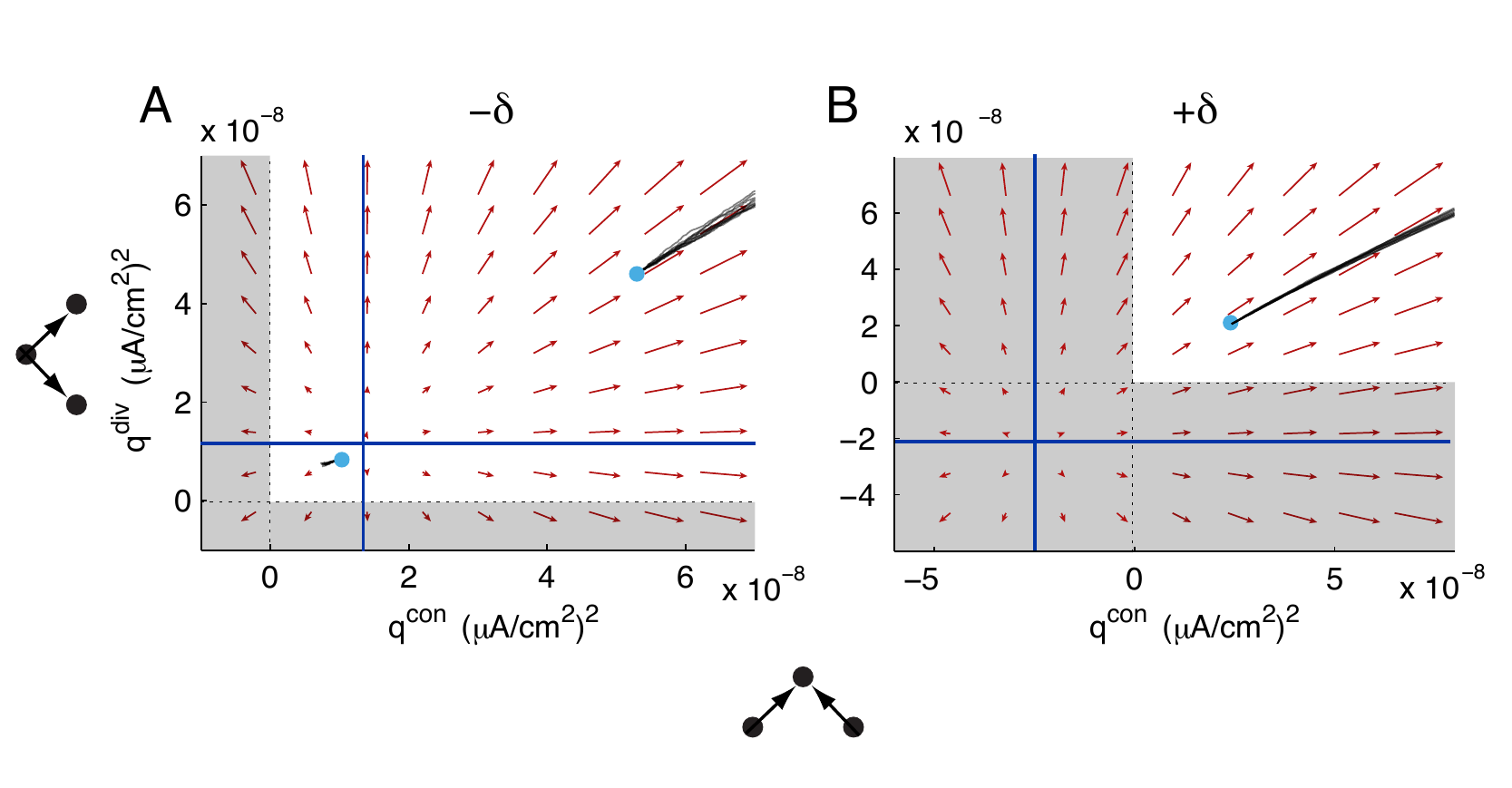} } }
\caption{
{\bf Plasticity of convergent and divergent motifs with balanced STDP.}  (A) Joint dynamics of convergent and divergent motifs when STDP is balanced and depression-dominated.  Initial conditions as in Fig. 7A.  (B) Joint dynamics of convergent and divergent motifs when STDP is balanced and potentiation-dominated.  Initial conditions as in Fig. 7B.  Red: the flow in the $q^\mathrm{con},q^\mathrm{div}$ slice of the motif phase space.  Black: plasticity of the motifs in simulations of the full spiking network.  Cyan dots mark initial conditions for the simulations and each line is a realization.  The vector fields are calculated with all other variables fixed at these initial conditions.  For (A), the vector fields are similar for both sets of initial conditions.  In both panels, blue lines mark projections of each variable's nullcline into the plane and regions of unattainable negative motif strengths are shaded.
}
\label{Fig8}
\end{figure}

%\newpage

\begin{figure}[ht!]
	{\center{\includegraphics{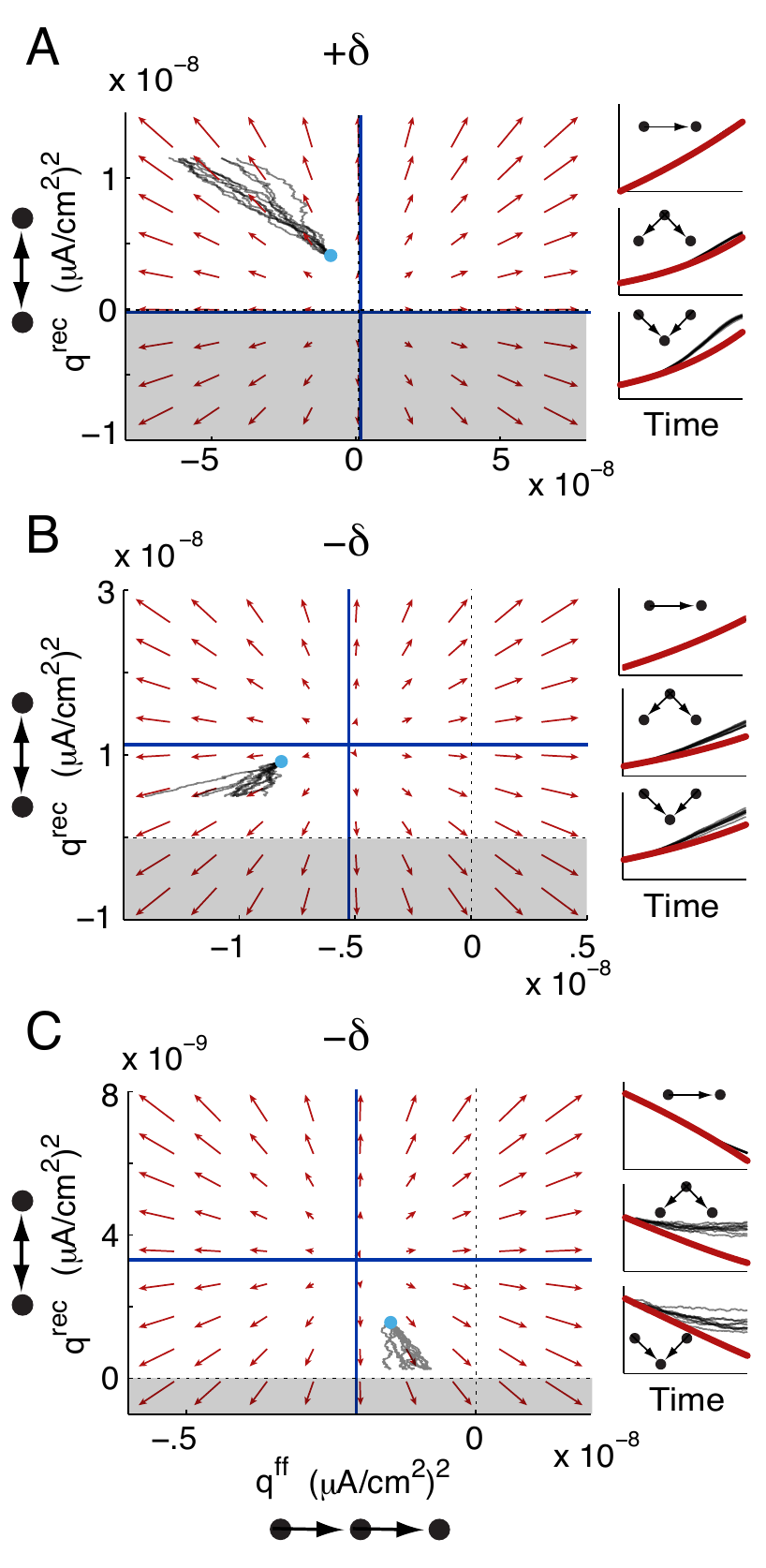} } }
\caption{
{\bf Plasticity of recurrent loops and feedforward chains with balanced STDP.}  (A-C) The dynamics of loops and feedforward chains, with all other variables fixed at their initial conditions.  In all cases, the projections of the $q^\mathrm{ff}$ and $q^\mathrm{rec}$ nullclines into this plane provide thresholds for the potentiation or depression of each motif.  The shape of the STDP rule and the initial values of the other motif variables determine the locations of these nullclines.  Color conventions are as in Fig. 8.  In each panel, right insets show the time series of $p$ (top), $q^\mathrm{div}$ (middle) and $q^\mathrm{con}$ (bottom), with spiking simulations in black and motif theory in red.  A) The potentiation-dominated balanced STDP rule.  B) The depression-dominated balanced STDP rule, in the region where $p$, $q^\mathrm{div}$ and $q^\mathrm{con}$ potentiate.  C) The depression-dominated balanced STDP rule, in the region where $p$, $q^\mathrm{div}$ and $q^\mathrm{con}$ depress.
}
\label{Fig9}
\end{figure}

%\newpage
%\section*{Supporting Information Legends}
%\subsection*{Balanced plasticity in isolated pairs of neurons}
%\begin{figure}[ht!]
%{\center{\includegraphics{supp/SuppFig1.pdf} } }
%\caption*{
%	{\bf Figure S1.  Both types of balanced STDP lead to splitting of synaptic weights.} We take an isolated pair of neurons with the same intrinsic and synaptic parameters as in the full network.  The neurons are reciprocally connected.  (A) The depression-dominated balanced STDP rule.  (B) The potentiation-dominated balanced STDP rule.  Both STDP rules are exactly as in the main paper.  In both cases, the reciprocal loop is eliminated and initial conditions determine which synaptic weight is potentiated and which depressed.}
%\end{figure}

%
% Please enter your Supporting Information captions below in the following format:
%\item {\bf}
%\item {\bf}
%\end{description}

\end{document}